\documentclass[manuscript,nonacm]{acmart} % arXiv version: de-anonymized, no ACM/venue metadata

\usepackage{booktabs}
\usepackage{graphicx}
\usepackage{tabularx}
\usepackage{array}
\usepackage{ragged2e}
\usepackage{longtable}
\usepackage{xcolor}
\newcolumntype{Y}{>{\RaggedRight\arraybackslash}X}
\newcolumntype{P}[1]{>{\RaggedRight\arraybackslash}p{#1}}

\definecolor{PreMiroQuote}{HTML}{245A81}
\definecolor{MiroQuote}{HTML}{754577}

\newcommand{\studyquote}[3]{%
  \textcolor{#1}{\textit{``#3''}~\textnormal{[#2]}}%
}
\newcommand{\premiroquote}[2]{%
  \studyquote{PreMiroQuote}{#1, pre-Miro interview}{#2}%
}
\newcommand{\miroquote}[2]{%
  \studyquote{MiroQuote}{#1, Miro activity}{#2}%
}
\newcommand{\premirofragment}[1]{%
  \textcolor{PreMiroQuote}{\textit{``#1''}}%
}
\newcommand{\mirofragment}[1]{%
  \textcolor{MiroQuote}{\textit{``#1''}}%
}
\AtBeginDocument{%
  }

\setcopyright{none}
\copyrightyear{2027}
\acmYear{2027}
\begin{document}

%%
%% The "title" command has an optional parameter,
%% allowing the author to define a "short title" to be used in page headers.
\title[Configuring AI for Exploratory Learning]{Configuring AI for Exploratory Learning: Educator Perspectives on Inference, Persistence, and Disclosure}

\author{Si Chen}
\affiliation{%
  \institution{University of Notre Dame}
  \city{Notre Dame}
  \state{Indiana}
  \country{USA}}
\author{Xinyue Chen}
\affiliation{%
  \institution{University of Michigan}
  \city{Ann Arbor}
  \state{Michigan}
  \country{USA}}
\author{Artur Mullagaliyev}
\affiliation{%
  \institution{University of Notre Dame}
  \city{Notre Dame}
  \state{Indiana}
  \country{USA}}
\author{Alexander Nwanganga}
\affiliation{%
  \institution{University of Notre Dame}
  \city{Notre Dame}
  \state{Indiana}
  \country{USA}}
\author{Shifu Hou}
\affiliation{%
  \institution{University of Notre Dame}
  \city{Notre Dame}
  \state{Indiana}
  \country{USA}}
\author{Deng Pan}
\affiliation{%
  \institution{University of Notre Dame}
  \city{Notre Dame}
  \state{Indiana}
  \country{USA}}
\author{Ronald Metoyer}
\affiliation{%
  \institution{University of Notre Dame}
  \city{Notre Dame}
  \state{Indiana}
  \country{USA}}
\author{Sugana Vijay Chawla}
\authornote{Corresponding author.}
\affiliation{%
  \institution{University of Notre Dame}
  \city{Notre Dame}
  \state{Indiana}
  \country{USA}}

\renewcommand{\shortauthors}{Chen et al.}

%%
%% The "author" command and its associated commands are used to define
%% the authors and their affiliations.
%% Of note is the shared affiliation of the first two authors, and the
%% "authornote" and "authornotemark" commands
%% used to denote shared contribution to the research.

%% The abstract is a short summary of the work to be presented in the
%% article.
\begin{abstract}
%Generative AI (GenAI) can support open-ended learning by generating experiences, interpreting learner activity, and adapting support. Yet educators often lack ways to understand how these capabilities work together or configure them around educational goals. We investigate this gap through staged design interviews with 15 U.S. educators around the topic of career exploration. Participants configured what AI could adapt, infer, remember, share, and decide, as well as when people should review or redirect it. They valued access and continuity but worried that temporary exploration could become a lasting judgment that narrowed future opportunities. They also distinguished AI that broadened exploration from more direct guidance that recommended what a student should pursue. We contribute an empirical understanding of educator configuration and three design directions for GenAI-supported learning: make AI inferences visible and contestable, limit memory and sharing to a stated purpose, and match human oversight to impact.
Generative AI (GenAI) can support open-ended learning through generation, personalization, and learner modeling, yet educators need ways to shape these capabilities around educational goals. Through interviews and design activities with 15 U.S. educators, we examined educator configuration of GenAI using K–12 career exploration as an exploratory context. Educators configured not only AI-generated experiences, but also when student activity became an inference, whether learner information persisted, who could access it, and how it informed subsequent human action. They also faced challenges translating teaching needs into configurations: recognizing possibilities for control beyond familiar uses of GenAI, decomposing general-purpose AI into understandable functions and responsibilities, and identifying useful information through intended teaching actions. We discuss how GenAI systems can support educators in expressing and testing configurations, while establishing boundaries around personalization, inference, persistence, disclosure, and action to keep AI-supported learning aligned with evolving learner needs.
\end{abstract}

%%
%% The code below is generated by the tool at http://dl.acm.org/ccs.cfm.
%% Please copy and paste the code instead of the example below.
%%


%%
%% Keywords. The author(s) should pick words that accurately describe
%% the work being presented. Separate the keywords with commas.
%% A "teaser" image appears between the author and affiliation
%% information and the body of the document, and typically spans the
%% page.
%%
%% This command processes the author and affiliation and title
%% information and builds the first part of the formatted document.
\maketitle

% Introduction included by main.tex.

\section{Introduction}
\label{sec:introduction}

Open-ended learning tasks, such as argumentative writing and inquiry-based
projects, require students to develop ideas without a single predetermined
solution \cite{kim2024Dashboard,srivastava2025LearnLens}. Supporting these
tasks requires guidance that responds to learners' changing needs and to work
whose meaning may emerge over time
\cite{hilliger2026Chatbot,do2025PAIGE}. Providing such support at scale is
challenging because educators must interpret individual progress and decide
how to guide subsequent learning
\cite{holstein2017TeachersAides,wiseJung2019TeachingAnalytics,prieto2018OrchestrationLoad}.
GenAI can generate activities, provide feedback, infer patterns from
interaction, and personalize subsequent support
\cite{bock2025LearnerModels,nguyen2024StudentModeling,hilliger2026Chatbot}.
Connecting these capabilities also creates dependencies: activity can become
an inference about a learner, the inference can shape later generation, and
generated information can reach an educator. Existing research has raised
concerns about inaccurate or biased learner representations and examined ways
for people to inspect or influence AI-supported learning
\cite{bullKay2016SMILI,conati2018Interpretable,williamson2020Historical,holstein2019Complementarity,lawrence2024SharedControl}.

Educators may respond to these concerns by avoiding GenAI or relying on
restrictions such as chatbot guardrails
\cite{han2024ElementaryLiteracy,dangol2026Relief,yan2024Challenges}. Such
restrictions do not by themselves specify a pedagogically useful role for AI.
Classroom orchestration research instead examines how educators coordinate
learners, activities, resources, and transitions, including when and how to
intervene \cite{dillenbourg2013Orchestration,roschelle2013Orchestration}.
Applying this perspective to GenAI raises two related design questions: how
educators can translate pedagogical intent into connected AI functions, and
how they can set boundaries on the learner information and human actions that
connect those functions over time.

Recent systems allow teachers to configure student--GenAI interaction, author
learning activities, and inspect pedagogical agents
\cite{ortegaArranz2025TeacherAutonomy,jin2025TeachTune,ko2026CourseTutors,li2026GenRole}.
These systems establish important forms of teacher control, while leaving open
how educators would configure relationships among several functions and the
later uses of information produced through them. We use \emph{educator
configuration} to refer to educators' ability to specify and revise GenAI
functions, their information dependencies and handoffs, and the boundaries on
how resulting learner information informs later support. This scope includes
educator control while recognizing that students and other educational roles
may need control at particular points.

We conducted an exploratory study to understand how educators would want to configure GenAI support for open-ended learning. To do this, we conducted interviews and staged design activities with 15 educators involved in K--12 career exploration.  %including teachers, a counselor, educational leaders, a workforce-development partner, and homeschooling parents.
Career exploration provides a setting for open-ended learning tasks in which students encounter possibilities and gradually develop a sense of what interests them. %Interests develop through experience and are not fixed traits \cite{hidiRenninger2006Interest}. 
%Adolescents also revise possible career identities through activities and conversations \cite{malanchuk2010CareerIdentity,marshall2008PossibleSelves}.
Supporting this process requires sustained guidance involving coordination among educators, families, and organizations beyond the school \cite{marshall2008PossibleSelves,malanchuk2010CareerIdentity,marcionetti2025Support}. Maintaining continuity is challenging, particularly when resources are limited \cite{marcionetti2025Support,damico2026PurposefulWork}. Career exploration therefore offers a demanding setting for examining how educators would translate complex educational needs into the design of AI-supported learning, and what control they would need over AI systems to keep that support aligned with educational goals over time.
%GenAI can broaden access to occupations and connect exploration to local
%courses or workplaces. However, the same system can turn a momentary choice
%into an inferred interest, store that inference as a profile, and use it to
%narrow later opportunities. Career support also crosses teachers, counselors,
%families, and institutions, making the timing and audience of AI-produced
%information as important as its apparent accuracy.
%To investigate these questions, we conducted interviews and staged design
%activities with 15 educators who support K--12 career exploration,
%including teachers, a counselor, educational leaders, a workforce-development
%partner, and a homeschooling parent. Participants responded to a concrete
%career exploration demonstration and arranged student-facing, teacher-facing,
%and backstage AI functions, information flows, controls, and handoffs in a
%Miro design probe. We treat the resulting boards as proposed designs created
%jointly during the study. They do not show how an implemented system would work
%or establish demand for a particular technical architecture.
We aim to answer three research questions through the study:

\begin{itemize}
  \item \textbf{RQ1:} What opportunities and concerns do educators identify
  for GenAI-supported K--12 career exploration?

  \item \textbf{RQ2:} How do educators envision configuring GenAI-supported
  career exploration?

  \item \textbf{RQ3:} What challenges and support needs emerge as educators translate their teaching needs and practices into GenAI configurations?
\end{itemize}

Addressing these questions required bridging educators' familiarity with
educational practice and their varied familiarity with AI. Through 15
formative sessions with 12 contributors, two of whom later joined the formal
study, we refined probes that made possible GenAI functions and relationships
available for discussion without requiring technical expertise. The formal
study with 15 educators began with participants' existing practices and needs,
followed by a video probe and a Miro-based design activity. Participants
selected, modified, or created AI functions for a learning scenario grounded
in their practice, connected these functions with students and educators, and
explained how information should move and when people should guide, review, or
stop AI support.
Participants saw value in using GenAI to broaden exploration, gather evidence
over time, and connect information across educational settings, while resisting
guidance that gave developing interests premature weight. Their configurations
specified what shaped a generated experience, when activity became an
inference about a learner, whether information persisted, who received it, and
what action followed. Across RQ3, we identified challenges and support needs that emerged as educators translated teaching needs and practices into envisioned configurations: moving from prevailing perceptions of AI toward concrete possibilities for educator control, turning general-purpose GenAI into configurable functions and responsibilities, and identifying useful information from open-ended interactions through intended teaching actions.

%Participants valued GenAI for broadening exploration, adapting access, and
%preparing human guidance. They also distinguished reversible adaptations from
%durable profiles, disclosures, interventions, and pathway recommendations.
%Their configurations located control at specific dependencies: which sources could inform an activity, when interaction became inference, what persisted, what crossed to an educator, and who could redirect the system.% The staged
%materials also helped participants move from discussing existing problems to
%specifying flows and stopping points. We treat this as a secondary
%methodological observation, separate from the three research questions.

This paper makes two main contributions. First, we provide empirical findings on educator configuration of GenAI-supported open-ended learning, including the opportunities and concerns educators identify, what they seek to configure across AI functions, learner information, and human involvement, and the challenges and support needs that emerge as they translate teaching needs into envisioned configurations. Second, we discuss implications for setting conditional and role-sensitive boundaries on configuration, supporting educators in translating teaching needs into inspectable configuration choices, and designing GenAI career exploration that supports evolving interests without prematurely narrowing possibilities. We additionally reflect on staged design probes for eliciting configurations while tracing how participants responded to and extended researcher-provided possibilities.

\section{Related Work}
\label{sec:related-work}
Research on teacher orchestration and human--AI control examines educators' roles in AI-supported learning, while work on AI for teaching and learning considers emerging capabilities and their use in practice. Career exploration provides the developmental context for our study, and participatory AI design informs our approach to eliciting educators' needs. Together, these areas motivate our examination of how educators can configure GenAI-supported learning around their educational goals and contexts.

\subsection{Educator Control of AI-Supported Learning}

Classroom orchestration coordinates learners, activities, resources, groups,
and transitions under practical constraints
\cite{dillenbourg2013Orchestration,roschelle2013Orchestration}. It requires
educators to notice relevant conditions, interpret them, and act with limited
time and attention
\cite{prieto2018OrchestrationLoad,wiseJung2019TeachingAnalytics}.
Dashboards can support awareness, but their value depends on whether the
information supports classroom decisions \cite{verbert2014Dashboards}.

AI changes both what educators coordinate and how work is distributed. Studies
of intelligent tutoring systems identify needs for timely information and
teacher-facing views of student activity
\cite{holstein2017TeachersAides,holstein2018ClassroomDashboard}. Research on
complementarity, hybrid adaptivity, and shared control distributes adaptation
across teachers, learners, and AI
\cite{holstein2019Complementarity,holstein2020HybridAdaptivity,molenaar2022Hybrid,lawrence2024SharedControl}.
Other work examines teacher-in-the-loop analysis, teacher-controlled activity
generation, and support for learning orchestration skills
\cite{cohn2024HumanLoop,walkington2026MultiAgent,chun2026ArguMath}. Together, these studies establish the importance of meaningful teacher involvement in AI-supported learning.

Co-orchestration research further shows that control is not a single system
property. Teacher and student control can both be high
\cite{eshelKohavi2003ClassroomControl}, while people and AI may divide tasks
or contribute to the same task \cite{holsteinOlsen2023CoOrchestration}.
Control preferences also vary with classroom conditions, stages, and decisions
\cite{echeverria2020DynamicTransitions,olsen2021SocialTransitions,yang2021PairingPreferences,yang2023PairUp,yang2026BalancingAgency}.
For example, teachers have preferred AI suggestions that they can inspect and
reject, while classroom studies have also found differences between teacher
and student preferences. These results suggest that control must be configured around both the decision and the people affected by it.
%Human involvement must therefore be designed for particular activities rather than assigned once for an entire system.
Human involvement must therefore be configured for particular activities and decisions rather than assigned once for an entire system.

GenAI brings this issue into a context where educators are still deciding
whether and how to adopt the technology. Research identifies potential
benefits for feedback and access alongside concerns about accuracy,
over-reliance, privacy, authorship, and age-appropriate use
\cite{kasneci2023ChatGPT,lo2023RapidReview,yan2024Challenges,han2024ElementaryLiteracy}.
Adoption is also shaped by professional values, institutional constraints,
training, infrastructure, and unequal access
\cite{dangol2026Relief,xiao2026Inequality}. Educators and technology providers may prioritize different concerns, including consent, lost context, bias, and unequal power \cite{harvey2025Teachers,han2026Sensing,han2025Helping}. These
conditions can make hesitation reasonable even when AI offers useful support.

Meanwhile, educational systems provide more ways to monitor, author, and
influence AI-supported activity. Teachers can review student--AI interaction,
configure model behavior, create learning activities, and inspect teaching
agents
\cite{kim2024Dashboard,srivastava2025LearnLens,ortegaArranz2025TeacherAutonomy,jin2025TeachTune,ko2026CourseTutors,li2026GenRole,echeverria2025TeamVision}.
Dashboards and alerts can support awareness during use, while authoring tools
and tutor-building systems give educators more influence before students begin
an activity.
These controls usually address one capability or moment in an activity. They
do not necessarily help educators understand how several capabilities work
together, anticipate their local effects, or change behavior that conflicts
with an educational purpose.

\subsection{GenAI Support for Teaching and Learning}

Technical support for teaching and learning is expanding quickly. Systems can
generate learning materials, sustain dialogue, provide feedback on developing
work, infer patterns from student activity, and recommend pedagogical support
\cite{do2025PAIGE,hilliger2026Chatbot,nguyen2024StudentModeling,dehbozorgi2024Recommender}.
Teacher-facing systems can summarize activity, support review, and help
educators create or adapt learning experiences
\cite{kim2024Dashboard,srivastava2025LearnLens,ortegaArranz2025TeacherAutonomy}.
Other pipelines connect and verify information from several sources
\cite{kwon2024Schema}. These developments increase the range of roles that AI
may take before, during, and after a learning activity.

Technical capability alone does not determine how a capability should be used
educationally. Studies have examined goal setting and rewards in tutoring,
purposeful support during writing, LLM-based formative assessment, and
different pedagogical agent roles
\cite{borchers2025GoalSetting,siddiqui2025Writing,dimitriadou2026Formative,xu2026Explain}.
Evaluations in mathematics education likewise show that possible learning
benefits must be considered together with failure risks
\cite{kumar2025MathEducation}. The same underlying capability can therefore
support different educational purposes and create different demands for
teacher understanding and involvement.

A learner model is one technical example, representing a student's knowledge,
characteristics, behavior, or needs for use in later support
\cite{bock2025LearnerModels}. Open learner models can make such a
representation available for inspection, editing, or discussion
\cite{bullKay2016SMILI,dimitrova2016Interactive,vanLabeke2007Interpretation}.
We return to this literature in the Discussion after learner representation
emerges from our findings.

The central gap is therefore not a lack of educational AI capabilities or a
general argument that a human should remain involved. It is the lack of ways
for educators to understand how capabilities are combined in a learning
experience and to translate educational intentions into system behavior they
can inspect and change. Work on interpretable and contestable AI and general
human--AI interaction provides foundations for understandable behavior and
ongoing control
\cite{conati2018Interpretable,alfrink2023Contestable,amershi2019Guidelines}.
However, educators still need ways to shape unfamiliar GenAI support without
having to specify a technical architecture. We investigate this gap by asking
educators to configure proposed AI support around the purposes and conditions
of their own practice.

\subsection{Career Exploration as Developmental Learning}

Career exploration continues over time. Interest develops through
interaction: triggered situational interest sometimes becomes a sustained
individual interest \cite{hidiRenninger2006Interest,renningerHidi2026Differences}.
Adolescents' ideas about future work change over time
\cite{malanchuk2010CareerIdentity}, and possible future selves develop through
family conversations and activities
\cite{marshall2008PossibleSelves}. This literature makes two design assumptions
unsafe. First, a momentary choice provides weak evidence of a stable trait.
Second, even a plausible recommendation can cause harm if it repeatedly narrows
what a young person encounters.

GenAI can nevertheless broaden access by translating technical information,
generating several formats, and making otherwise unavailable experiences more
concrete. HCI and AI-in-education studies report benefits from personalized generated
podcasts and student-designed learning applications that use several media
\cite{do2025PAIGE,prasad2025Multimodal}. Teacher-driven context
personalization likewise illustrates how educators can connect generated
problems to student interests \cite{walkington2025TeacherDriven}. At the same time, students who place
high trust in GenAI may report less critical thinking \cite{lee2025CriticalThinking}, and youth
need scaffolds for recognizing hallucinations and other limitations
\cite{tian2026Hallucination}. Career exploration therefore provides a demanding
case for educator configuration: the system should increase meaningful
encounters while
keeping interpretations open to revision and connecting generated experiences to
reflection, local opportunity, and human guidance.

\subsection{Participatory Approaches in AI Design}

Participatory AI design involves people affected by a system in shaping its
purposes, data, roles, and controls. Yet participation is always structured by
the materials, examples, and roles that researchers provide
\cite{sandersStappers2008Cocreation}. This is especially important for AI.
Domain experts may understand the practice being supported while having
different levels of familiarity with AI capabilities and limitations
\cite{longMagerko2020AILiteracy}. Design materials must therefore make
unfamiliar or hidden system relationships discussable without treating
researcher-provided examples as the only possible design.

Participants also need resources for reasoning about AI, not only opportunities
to react to a proposed system. Research on children's AI learning shows that
content, presentation, and activity design affect what people can understand
\cite{jia2025ChildrenAILearning}, while young people's interpretations of AI
are shaped by their social contexts \cite{dai2026EthicalReasoning}.
Educator-facing work has used structured reflection to help teachers question AI behavior
\cite{anthis2025Heuristic}. Research on student agency further distinguishes
ways that learners can choose, direct, or decline AI involvement
\cite{vincoli2025Agency,retes2026Agency}. Together, this work frames
participation as supporting both design input and the ability to make informed
choices about AI.

Prior studies use concrete artifacts to support this process. Data probes make
data relationships and their implications available for critique
\cite{zhang2023DataProbes}. Game-based co-design can surface how young people
understand AI \cite{lim2025Escape}, while multimodal GenAI workshops help
participants explore possible roles and applications
\cite{prasad2025Multimodal}. Participatory speed dating supports comparison
across predefined control arrangements \cite{yang2026BalancingAgency}. These
approaches make unfamiliar systems easier to discuss, but the artifacts also
shape which ideas become salient. Speed dating supports comparison across
alternatives, for example, while a reconfigurable artifact can help
participants construct relationships that researchers did not predefine.

Our staged approach addresses this tension by combining unaided discussion, a
researcher-authored video, and a reconfigurable Miro activity. Participants
first described their existing practice, then examined a shared example, and
finally rearranged, removed, or created system functions and connections.
Tracking when ideas appeared allowed us to distinguish participants' initial
concerns from responses to our example and configurations developed through
the design activity. This approach enabled educators to specify how they
wanted to shape AI support without requiring them to describe a technical
architecture.

\section{Method}
We conducted one-on-one interviews with 15 U.S. educators involved in K--12 career exploration. The 15 interviews used design probes that we iteratively refined through additional  three rounds of formative sessions with 12 contributors; two of these contributors later participated in the formal study. Our study examined how educators envisioned GenAI-supported career exploration, focusing on four aspects of system design: what educators should be able to configure, how they imagined learners interacting with and experiencing an AI-supported career-exploration system, what information should be shared across activities and stakeholders, and when human interpretation or intervention should be required.

\begin{figure*}[t]
    \centering
    \includegraphics[width=\textwidth]{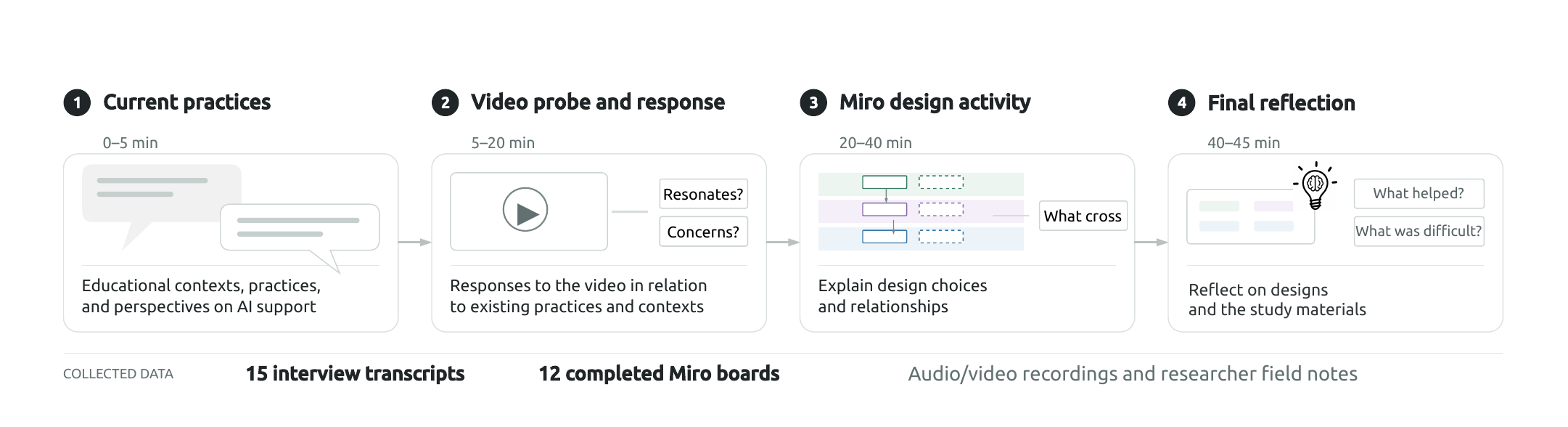}
    \caption{Overview of the interview process. Participants discussed their current practices and initial ideas, viewed and reflected on a video probe, and developed and walked through their proposed designs in Miro. The interview concluded with a reflection on their designs and the study materials.}
    \Description{A four-stage interview sequence: initial discussion of current practices and ideas; video viewing and reflection; Miro design activity and walkthrough; and final reflection.}
    \label{fig:study-process}
\end{figure*}

Each interview consisted of three main activities (Figure~\ref{fig:study-process}): (1) an initial exploration of participants' current practices and perspectives on AI-supported career exploration, (2) a video probe followed by participants' responses, and (3) a think-aloud design activity conducted in Miro. We concluded with a brief reflection on participants' designs and their experiences with the study materials. We describe each activity in detail below. The study received approval from the university's institutional review board.

\begin{itemize}
\item \textbf{Exploring current practices and perspectives (0--5 minutes).}
Before introducing the design probes, we asked participants about their educational contexts, current career-exploration practices, constraints, experiences with AI, and ideas for how AI might support career exploration. This was designed to capture participants' initial ideas before they saw researcher-provided examples.

\item \textbf{Video probe and response (5--20 minutes).}
Participants then viewed an approximately five-minute video probe depicting possible GenAI-supported interactions for career exploration. We then asked what resonated with them, what raised concerns, and how the depicted features might fit with or conflict with their existing practices and contexts.

\item \textbf{Miro design activity and walkthrough (20--40 minutes).}
Participants were then asked to select, combine, or create a career-exploration scenario and used the Miro board to design a system they would want to use. They were asked to think aloud while they created, %Thinking aloud, they selected, modified, combined, or created AI functions and positioned them in relation to students, teachers, and backstage processes. 
explaining their choices and the relationships among the elements as they developed their designs.

\item \textbf{Final reflection (40--45 minutes).}
Finally, we asked participants to reflect on what their configurations made possible or difficult, how they understood the relationships they had constructed, and how the video probe and Miro materials supported or constrained their thinking.
\end{itemize}

We recruited 15 participants (T1--T15), including classroom teachers, a school counselor, educational and career-pathway leaders, and homeschooling parents (Table~\ref{tab:participants}). Participants represented elementary, middle, and high school contexts. All participants worked in the same city in the Midwestern United States. We recruited participants through local school and workforce-development partnerships, contacts from prior studies, and participant referrals. Participants received \$$25$ or \$$30$ USD in compensation (depending on session length) and participated in sessions lasting approximately 45--60 minutes. Throughout the paper, we use \emph{educator} as an umbrella term for all participants, as each supported educational activities in a school, community, or home setting. We refer to participants by their specific roles, such as teacher or counselor, when those distinctions are relevant.

We use \emph{student} for a young person in the envisioned K--12 setting and
\emph{learner} when discussing learner models or broader design implications.

%Educator roles carry different
% responsibilities and forms of control. In this paper, students can inspect and
% contest representations of themselves and decide whether learner-controlled
% records transfer across sessions. Teachers control classroom activities,
% instructional materials, source choices,
% and session-level intervention. Counselors and pathway staff review
% consequential pathway guidance and disclosures needed for that support, while
% school or district leaders establish institutional rules for deployment,
% access, and retention. These responsibilities can overlap locally; GenAI may
% execute a delegated action, but it does not independently decide what should
% happen to a learner.

\begin{table*}[t]
\centering
\caption{Participant roles and educational contexts.}
\label{tab:participants}
\small
\begin{tabularx}{\textwidth}{P{0.08\textwidth} Y P{0.24\textwidth}}
\toprule
\textbf{ID} & \textbf{Educational role/context} & \textbf{Grade} \\
\midrule
T1  & Science teacher & High school \\
T2  & Science teacher & Grade 8 / secondary \\
T3  & Teacher/advisory context & Middle school \\
T4  & School counselor & Grades 10--12 \\
T5  & Workforce and career-exploration partner & Cross-school \\
T6  & Middle-school college/career teaching context & Grades 7--8 \\
T7  & Principal & Secondary \\
T8  & Pathways/curriculum coach & K--12 \\
T9  & School leadership/advisory context & Middle school \\
T10 & District CTE director / AI task-force lead & K--12 \\
T11 & Former teacher / homeschooling parent, primary instructor, and tutor & K--8 \\
T12 & Homeschooling parent / primary instructor & Grades 7--12 \\
T13 & STEM teacher & K--6 \\
T14 & Engineering/technology teacher & High school \\
T15 & Industrial technology teacher & Grade 9 / high school \\
\bottomrule
\end{tabularx}
\Description{Fifteen participants are listed by anonymized ID, educational
role or context, and grade band.}
\end{table*}

\subsection{Iterative Development of the Interview Process and Design Probes}

We aimed to understand educators’ needs for AI-supported career exploration and how they would want to shape that support. Doing so required establishing common ground for discussing how AI capabilities might relate to their educational goals. We iteratively developed the interview process and study materials to build this shared understanding and support educators in expressing their expectations.

To achieve this goal, we conducted 15 formative study sessions with 12 distinct contributors between January and August 2026. Three educators participated in two sessions, and one contributor participated through a written response. The formative group included school-based educators across elementary, middle, and high school settings, as well as a local workforce-development partner. The participating teachers worked in the same Midwestern U.S. city. Two contributors later participated in the formal study.

\begin{figure*}[h]
    \centering
    \includegraphics[width=\textwidth]{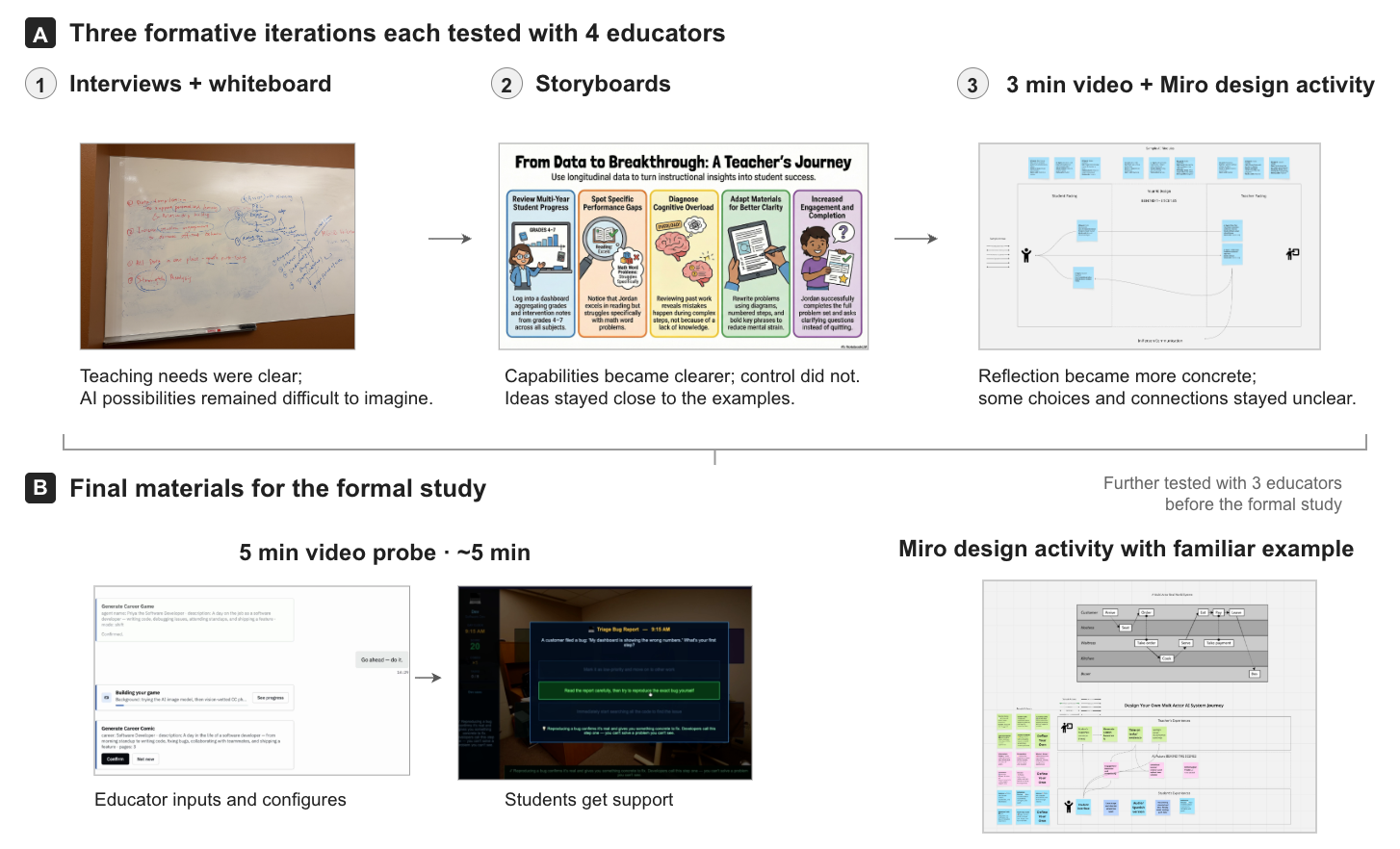}
    \caption{Iterative development of the study materials.
    (A) Three formative iterations informed successive revisions
    to support educators in understanding AI capabilities and
    expressing their design expectations.
    (B) The final video probe and Miro design activity were further
    tested with three educators before use in the formal study.}
    \label{fig:probe-development}
\end{figure*}

The formative study proceeded through three main iterations (Figure~\ref{fig:probe-development}). In the first iteration, we used broad prompts and a whiteboard activity asking 4 educators what GenAI features might support their students in career exploration and how they would like to engage in the process. These sessions surfaced educators' interest in supporting career exploration longitudinally. However, these sessions revealed a gap between the questions we hoped to explore around how educators would like to configure AI-mediated learning experiences and what participants could readily articulate. Educators could explain their teaching goals and students’ needs, but their understanding of AI capabilities was often limited or centered on a narrow range of familiar uses. This made it difficult to discuss what support they might want AI to provide or how they would want to configure it. We therefore recognized a need to first familiarize participants with relevant AI possibilities, establishing a shared basis for a more concrete discussion of their needs.

%, but participants often discussed AI at a high level and found it difficult to envision what it might concretely do. This motivated us to introduce more concrete representations of possible AI capabilities in the next iteration.

In the second iteration, we developed storyboards illustrating how AI could address educational needs raised in the initial sessions. These examples helped participants develop a more concrete understanding of possible AI capabilities. However, seeing what AI could do did not necessarily clarify what educators could change or decide, or how they could respond if its support became inappropriate or potentially harmful. For example, a storyboard showing personalized guidance was preferred by educators, but they were confused about what information shaped that guidance or how they could influence it. In a formative session with 4 contributors, we observed that in describing their ideas, educators often spoke about AI as if different people were providing different kinds of support; for example, someone helping students explore and someone helping teachers guide that exploration. This led us to consider breaking down AI functionality into familiar concepts, such as different roles, actors, and activities, to help educators understand and discuss how different forms of support could work together.  At the same time, participants’ ideas often remained close to the storyboards, suggesting a need for materials that would help them develop alternatives beyond the examples provided.
%In the second iteration, we developed researcher-created storyboards that translated and expanded on ideas from the initial sessions. The storyboards helped participants consider a broader range of AI capabilities and interactions, but also surfaced two challenges. First, participants repeatedly distinguished between support for students and support for teachers, while finding it difficult to decompose ``AI'' into distinct actors with different roles and relationships. Second, although concrete examples helped participants envision possibilities they had not initially considered, the examples could also anchor their thinking to the capabilities and interactions depicted in the storyboards. These observations motivated us to make both the different AI actors and participants' ability to respond critically to researcher-provided examples more explicit in the next iteration.

Building on these observations, the third iteration introduced two complementary materials. 
First, we presented an approximately three-minute video illustrating an AI-supported career-exploration experience. The video presented a teacher-facing AI assistant for creating and configuring activities and a student-facing interface for interacting with AI career practitioners, such as a nurse or electrician, through career conversations and job simulations. Second, we developed a Miro design activity in which educators arranged role and activity cards to express what they wanted to configure, what information AI should process, and how they envisioned students interacting with the resulting support. This gave participants a way to express their expectations through proposed activities and relationships without having to specify a technical implementation. In formative sessions with four contributors, this combination helped participants recognize unfamiliar AI capabilities, reflect on them, and connect them to their own practice.

Two difficulties remained from the third iteration. Although the video showed a teacher-facing interface, participants needed more explicit examples of what teachers could change. Some participants also found it difficult to arrange roles and activities in Miro without a familiar example of how different parts could work together. For the final materials used in the formal study, we expanded the video to approximately five minutes, adding clearer connections between educator inputs and student-facing support. We also introduced a restaurant example before the Miro activity, showing how different roles, like customers, servers, and kitchen staff perform connected activities. Further testing with three educators indicated that the final materials provided sufficient support for eliciting their design expectations. We therefore adopted this version for the formal study. The following sections describe the final video probe and Miro design board in detail.

These rounds of iteration were intended to support the elicitation of design expectations from educators with varied familiarity with AI. The final materials provided concrete detail about possible AI support and how participants could shape it while leaving room for their own pedagogical goals and alternatives. The video established a shared reference for reflection, while the Miro activity enabled participants to develop proposals grounded in their educational practice. We used the two materials together so that the study extended beyond feedback on a predefined system.

\subsection{Video Probe }
\label{sec:video-demonstrations}
\begin{figure*}[h]
    \centering
    \includegraphics[width=\textwidth]{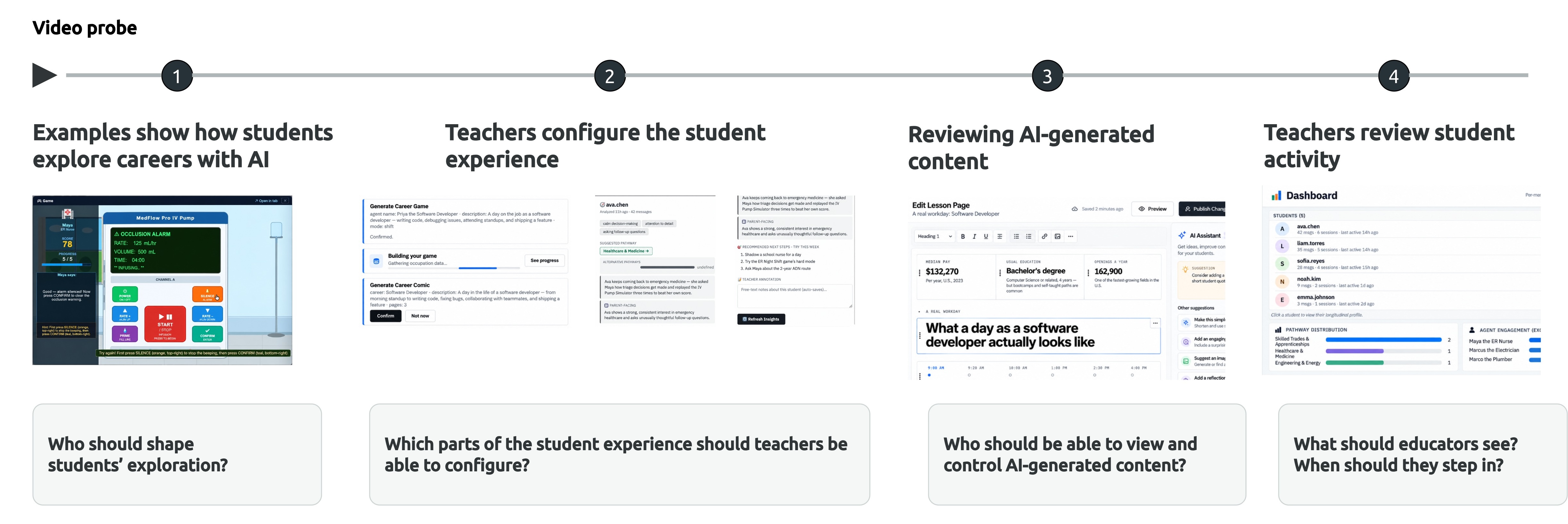}
   \caption{The five-minute video probe presented possibilities for how
educators could configure AI-supported career exploration, including
shaping what students could see and how educators can configure and control the students' activities.
Examples of student experiences and teacher-facing capabilities were
paired with questions to prompt educators' thoughts.}
    \label{fig:video-probe}
\end{figure*}

We used a five-minute video probe to provide a shared point of reference for discussing possible GenAI-supported career-exploration interactions (Figure  \ref{fig:video-probe}). Formative piloting suggested that verbal descriptions alone left the design space too abstract. We introduced the video after the initial open-ended discussion so participants could first describe their own practices and ideas before encountering a researcher-created example.

The video alternated between examples of user interface and questions about the design choices they raised. It began with a student-facing career-exploration experience and then introduced teacher-facing tools for creating and adapting that experience. It prompted participants to consider who should shape students' exploration, what teachers should be able to see, when teachers or counselors should become involved, how AI-generated experiences should be reviewed and controlled, and how career content might be presented through different forms of interaction. Together, these examples made student- and teacher-facing AI roles and their relationships more explicit.

The video served as a starting point for critique and reflection, not as a complete or ideal system design. Questions interspersed throughout the video highlighted assumptions and tradeoffs in the depicted interactions. After viewing it, participants discussed what resonated, what concerned them, and what they would change for their own contexts.

\subsection{Miro Design Probe}
\label{sec:miro-probe}

\begin{figure*}[h]
    \centering
    \includegraphics[width=\textwidth]{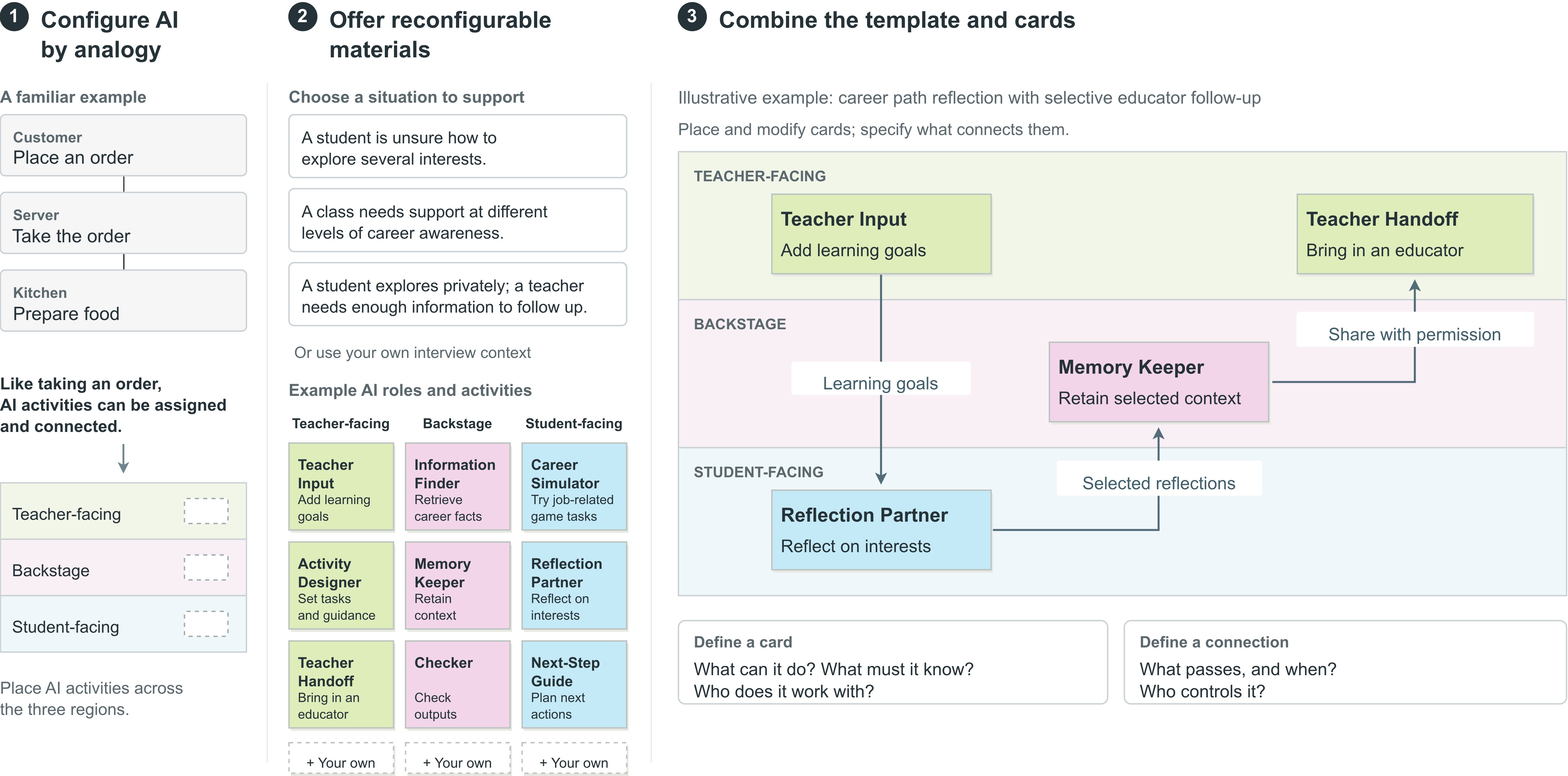}
    \caption{The Miro design probe. (1) A three-panel template supported configuration, with a restaurant analogy illustrating how to assign and connect activities. (2) Career-exploration scenarios and editable example cards supported participants in imagining AI roles and activities. (3) An illustrative arrangement showed how participants could combine and modify these materials to specify activities, connections, and control. The arrangement was a researcher-provided example, not a participant design.}
    \Description{Three panels explain the Miro design activity. The first uses customers, servers, and kitchen staff as an analogy for connected activities. The second provides career-exploration situations and editable cards for teacher-facing, backstage, and student-facing AI roles. The third shows a researcher-provided example connecting teacher input and handoff, backstage memory, and student reflection.}
    \label{fig:miro-probe}
\end{figure*}

The Miro design probe translated interactions from the video into materials participants could reconfigure (Figure~\ref{fig:miro-probe}). Participants could continue with a career-exploration scenario the came up with earlier in the interview or choose from three provided scenarios: a student navigating multiple interests, a teacher supporting students with different levels of career awareness, or balancing independent AI-supported exploration with educator awareness and follow-up. The scenarios provided context without prescribing a grade level or outcome.

The board was informed by Coordination Theory, which defines coordination as managing dependencies among activities \cite{maloneCrowston1994Coordination}. For example, in a restaurant, actors such as customers, servers, and kitchen staff perform different but interdependent activities, requiring coordination among them. We applied this framing to career exploration by asking which actors should perform different activities, how their activities depend on one another, and how those dependencies should be managed. Here, \emph{actors} could include students, educators, and distinct AI systems or agents, while activities included retrieving career information, supporting reflection, retaining context, checking outputs, or summarizing activity.

The board organized these activities into three panels: \textbf{student-facing}, \textbf{teacher-facing}, and \textbf{backstage}. The student-facing panel represented activities and AI actors that students would directly encounter. The teacher-facing panel represented those involved in configuration, review, and follow-up. The backstage panel represented activities such as retrieval, memory, routing, and output checking that might not be directly visible to either group. This structure helped participants reason about which actors should perform different activities, where those activities should occur, and what should remain visible or hidden. Moving a card between panels expressed a change in visibility or responsibility, while connections represented dependencies among actors and activities.

Each panel included example cards, blank cards, and connectors. Participants could keep, rename, modify, combine, separate, move, or remove the examples and add their own. Because formative sessions indicated that many educators were unfamiliar with Miro, the researcher shared the board on screen and manipulated it at the participant's direction. Participants specified which cards to use or change, where to place them, what to add, and how elements should connect.

As the board developed, we asked which actor should perform each activity, who should encounter it, what it needed from other parts of the system, and which activities should operate together or separately. We also probed what information educators should know, who should control different decisions, and when a teacher, counselor, or other person should become involved. Participants specified what should pass between connected elements, such as information, requests, permissions, feedback, or notifications.

The activity ended with a walkthrough of the completed Miro board. Participants explained how students and educators would experience the system, how actors and their activities would work together, and the reasoning behind their design choices.

\subsection{Data Collection and Analysis}
\label{sec:data-analysis}
We collected audio/video recordings, 15 interview transcripts, 12 completed Miro boards, and researcher field notes. Sessions were transcribed and de-identified before analysis. Three interviews did not produce a completed Miro board. In two of these interviews, detailed verbal responses filled the available time, so we did not proceed to the final design activity. Board completion was not required for inclusion in the interview analysis. Analyses of the design artifacts draw on the 12 completed boards.

We began with iterative, open-ended coding of the interview transcripts. Three
researchers independently reviewed the transcripts and revisited session
recordings as needed for context. Each researcher independently developed initial codes based on recurring patterns in the data. After analyzing eight interviews, the researchers met to compare codes, interpretations, disagreements, and areas requiring further evidence. We consolidated the codes into a shared codebook and grouped related
codes into preliminary themes. After analyzing three additional interviews, we met again
to refine the codebook and themes as new patterns emerged. For each code, we
recorded a definition and supporting transcript excerpts. As data collection
continued, we applied and refined the codebook and tracked patterns across
participants (see Appendix~\ref{app:codebook} for the final thematic codebook).

We also examined when ideas became visible across the interview stages. We
separated excerpts from the unaided interview, response to the video, and Miro
activity using the protocol sequence and transcript timestamps. In the unaided
interview, we focused on practices, problems, constraints, and initial ideas
articulated before participants encountered the probes. In the video response,
we examined what participants recognized, critiqued, questioned, or connected
to their own practices. In the Miro artifact and its accompanying think-aloud
and walkthrough, we examined proposed AI functions, information flows, forms
of control, and points of human involvement. In the final reflection, we examined participants'
comments on their designs and on how the study materials supported or
constrained their thinking. We did not code the researcher-authored video as
participant data. We coded what participants said in response to it.

For each central finding, we recorded the earliest stage in which related
evidence appeared in the collected data and whether the video or Miro starter
materials had already introduced the relevant capability or relationship. This
is a provenance marker rather than a causal claim. The initial interview did
not ask about every capability later shown in the probes, so absence before the
video does not show that the probe produced an idea. Appendix
Table~\ref{tab:finding-provenance} reports this comparison.

After forming the preliminary transcript themes, we open-coded each completed
Miro board using an artifact-focused codebook. We recorded whether a card was
provided, modified, removed, or created by the participant, where it was placed
across the student-facing, teacher-facing, and backstage areas, and how arrows
represented information flows, control, handoffs, stopping points, or
connections across time. The artifact codebook therefore included card source
and modification, panel placement, information flow, control point, handoff,
stopping point, and temporal connection. Across boards, these codes helped us
examine transcript themes such as educator-managed information sources,
learner-mediated memory transfer, visible GenAI inference histories, and return to
human guidance.

We compared each completed board with the starter materials. Retaining a
provided card indicated recognition or use of the example, not independent
generation of that function. We treated a board as elaborating the material
when a participant renamed, moved, combined, rejected, or replaced a provided
card, created a new card, or defined a new relationship and explained that
change during the think-aloud or walkthrough. This comparison allowed us to
distinguish the presence of a supplied function such as memory or handoff from
the participant's specification of what it retained, when it operated, what it
shared, or whether it should exist at all.

We treated the transcripts as the primary evidence for the themes and used the boards to complement the transcript analysis by showing how participants arranged functions and relationships. We did not interpret a sticky note, position, or arrow on its
own. Each board was analyzed alongside the participant's think-aloud comments
and final walkthrough, which explained what an element meant and why it was
connected to another. We then compared the board codes with the transcript
themes to elaborate, qualify, or provide contrasting configurations within
those themes rather than generating a separate set of findings.

% Drafting note: participant IDs were reconciled with the September 2 tracker.
% Before submission, verify every quotation against its audio recording and
% reconcile the research-question wording with the Introduction.
\section{Findings}
\label{sec:findings}

We organize the findings around three questions. We first establish what participants saw as useful and risky in GenAI-supported career exploration (RQ1). We then examine how they envisioned configuring GenAI-supported career
exploration and what they'd like to configure
(RQ2). Finally, we examine how educators moved from broad expectations of GenAI toward configurations grounded in teaching practice (RQ3).
% Throughout,
% we use design artifact to mean a Miro board assembled by a participant
% with facilitator-provided and participant-authored elements. These artifacts
% were created jointly during the activity and they show configurations participants
% considered. They do not show how an implemented system would work or how common
% a preference is. 
Quotation labels distinguish the pre-Miro interview, including responses to the video, from the Miro activity, including the think-aloud, walkthrough, and
immediate reflection.
% Counts report how many participants supported a claim in
% the 15 completed interviews or 12 completed design artifacts. We count each
% participant once for each claim. The counts are not population estimates, rather, they show how evidence was spread across this deliberately selected qualitative sample.

\subsection{RQ1: Opportunities and Concerns Educators Identify for GenAI-Supported Career Exploration}
\label{sec:findings-rq1}
%Career exploration is open-ended, unfolds over time, and connects students with local resources and human relationships. %Participants considered how GenAI could
%broaden exploration and connect interests to practical next steps, but also how
%it might give tentative interests premature weight. The three themes examine
%the breadth, pacing, and aggregation of GenAI-supported career exploration.
In this section, we examined participants’ existing career-exploration practices and how they envisioned GenAI supporting this work. They highlighted GenAI's opportunities to broaden students’ exposure, sustain support over time, and connect information across educational settings. These opportunities also raised concerns about narrowing students’ options, accelerating decisions, and losing context when combining observations. We organize these opportunities and associated tradeoffs into three themes below.

\subsubsection{Expanding broad possibilities without narrowing exploration}
\label{sec:rq1-open-exploration}

Four participants (T3, T10, T11, and T13) described early career learning as
exposure, play, and reflection before selection. Students often knew only the
occupations represented in their families, schools, or immediate communities.
T13, for example, said that students should encounter careers they did not yet
know existed. T11 similarly explained that students, parents, and teachers were
limited by the occupations they already knew.

Participants saw GenAI as a way to extend this existing work. They imagined it could make a
wider range of occupations available for students to consider. T11 wanted AI
to suggest possibilities beyond those that a student, parent, or teacher could
readily name. She described its intended role as
\miroquote{T11}{expanding our options}.
% TODO: verify quotation against audio

Participants were concerned that GenAI might narrow students’ exploration by steering them toward particular careers too early. This concern reflected a broader priority in early career education. T10, a district career
and technical education leader, emphasized that early career learning should involve  \premirofragment{not pigeonholing [kindergartners or third graders]
into a career} but instead \premiroquote{T10}{exploring and reflecting}. T11
made a similar distinction for GenAI. She wanted it to offer multiple
possibilities that students could explore and reconsider.  These accounts point to a need for GenAI support that keeps diverse career options open as students develop their interests. %identify
%scope as the first concern. GenAI broadened exploration when it added
%possibilities and narrowed exploration when it ranked or removed them.

\subsubsection{Gathering evidence without accelerating decisions}
\label{sec:rq1-developmental-staging}

%Providing different forms of career support as students develop was already
%part of participants' educational practice. 
Participants described different forms of career-exploration support as appropriate at different stages of students' development. Their existing practices ranged from early exposure and reflection to more focused course planning and pathway decisions. Schools also set timelines for these decisions. T7 reported that students created a college and career plan in
seventh grade and, by February of eighth grade, had to choose two high-school
pathways. She saw a role for earlier support because
\premiroquote{T7}{high school is too late}, while also observing that
\premiroquote{T7}{we're pressing them to make a decision}.
This highlighted the need to adjust the pacing of GenAI support to students' stages of exploration, particularly when to move from observing interests to offering explicit career guidance. %T13 similarly
%recalled that choosing a pathway \premiroquote{T13}{felt like a lot of pressure
%on her to have to choose a pathway as a freshman---really as an eighth grader
%going into high school}. 
%These institutional timelines created demand for
%support before students reached high school.
% TODO: verify T7 and T13 quotations against audio.

T12 made the difference in pacing explicit. In his current practice, he noticed his son's sustained interest in technical and STEM activities, but explained
that \premiroquote{T12}{it's not that we're saying, hey, you're going to become a computer scientist, or hey, you're going to become an engineer}. The family
continued to observe what interested him without organizing his experiences around a specific career. After viewing the GenAI experience, T12 called its
approach \premirofragment{very direct} because
\premiroquote{T12}{you're already talking about careers}, while in his view, most seventh- and eighth-grade students did not yet spend much time thinking about them. He
nevertheless thought AI could help parents and teachers notice patterns they might miss and \premiroquote{T12}{anticipate something you can't see directly}.
% TODO: verify T12 quotations against audio.

Participants also described more focused guidance when students were preparing
to act. T4 wanted AI to connect a student's interests with courses available in
the school, then use those options in a conversation about a four-year plan.
These examples distinguish pacing from scope. Interests became visible through
activity and observation over time. AI could help gather evidence during development,
but treating an early pattern as a direction for action could intensify the
existing pressure to decide.

\subsubsection{Aggregating fragmented information without collapsing context}
\label{sec:rq1-continuity}

The third theme concerned observations of a student's exploration that were
fragmented across subjects, teachers, grade levels, and the transition from
middle to high school. T7 described college and career work in her middle
school as \premiroquote{T7}{distributed through many people}. No single person
or setting necessarily held the observations that accumulated around a
student.
% TODO: verify T7 quotation against audio.

Participants saw GenAI as a way to aggregate these observations. T11 noted that
high-school students had \miroquote{T11}{multiple teachers a year}, a pattern
that compounded over four years. She contrasted a student finding algebra
\mirofragment{kind of bumpy} with geometry being
\mirofragment{his jam}, showing how observations could differ across
subjects. T11 wanted such observations to contribute to an
understanding that could be updated across teachers and years. T10 similarly
described a journal that followed students across grade levels so they could
revisit earlier exploration and \premiroquote{T10}{add on to it}. The
opportunity was to combine observations that were otherwise dispersed.
% TODO: verify T10 and T11 quotations against audio.

T11 also placed limits on aggregation. She wanted the process to remain
\mirofragment{primarily student-driven}. Observations
tied to a particular subject, teacher, or year could lose that context when
combined. The resulting collection could also contain more student information
than any one setting previously held.
% TODO: verify T11 quotation against audio.

\subsection{RQ2: Envisioned Configurations for GenAI-Supported Career Exploration}
\label{sec:findings-rq2}

% Across interviews and design artifacts, participants configured GenAI by
% deciding what context it received, how it interpreted student activity, and
% how its output could be used afterward. We relate these configurations to the
% three concerns in RQ1. The scope of
% exploration became a question of what GenAI personalized and how directly it
% guided students. Development over time became a question of when observations
% could become inferences and which information persisted. Support across people
% and settings became a question of what information reached whom and what they
% could do with it. The findings first orient the reader to the configurations
% participants produced and then examine these choices in turn.

Across interviews and design artifacts, we examined what educators wanted to configure in response to the opportunities and concerns identified in RQ1. Keeping exploration diverse raised questions about what GenAI should personalize and how directly it should guide students. Adapting support to students' readiness raised questions about when GenAI should observe, infer, or offer more focused guidance. Connecting support across people, settings, and time raised questions about what information should persist, who should receive it, and how they could use it. The following sections examine these configuration needs and the educational considerations behind them.

% Figure~\ref{fig:illustrative-codesign-artifacts} provides a visual guide to
% three configurations discussed across RQ2. Each column pairs the participant's
% original Miro board at the top with a simplified diagram created by the
% research team. The diagrams preserve the teacher-facing, backstage, and student-facing areas
% and the main connections among them. The T10 column shows
% educator-selected sources shaping a student experience. The T14 column
% connects student activity, a record carried between sessions, and teacher
% planning. The T15 column shows when GenAI makes an inference, what information
% is visible, and when a teacher can redirect the activity.

\begin{figure*}[t]
  \centering
  \includegraphics[width=\textwidth]{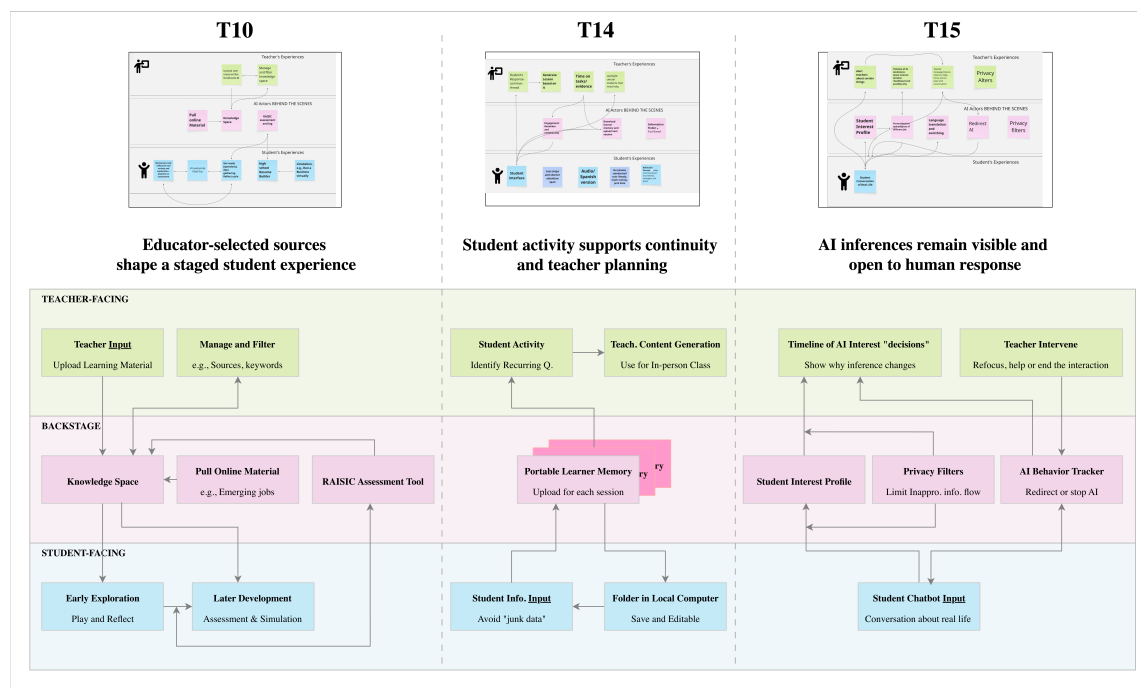}
  \caption{Original Miro design artifacts and simplified diagrams created by
  the research team for T10, T14, and T15. The original boards appear at the top. The diagrams
  below retain the teacher-facing (green), backstage (pink), and student-facing
  (blue) areas and the main relationships among selected cards. We condensed
  and lightly edited card wording for readability using the corresponding
  board, think-aloud explanation, and final reflection. Start reading from “Input” for each chunk and follow the arrows. Bidirectional arrows can be followed in either direction.}
  \Description{Three columns pair original Miro boards with simplified
  diagrams created by the research team. T10 connects teacher-selected learning materials
  and rules about information sources to backstage information and student
  activities. T14 connects student input and a learner record stored on the
  device to the teacher's use of GenAI to generate content. T15 connects
  student input to a chatbot and an interest profile to privacy controls, a
  timeline showing when GenAI inferred interests, and a teacher's decision to
  intervene. Green areas are teacher-facing, pink areas are backstage, and
  blue areas are student-facing. }
  \label{fig:illustrative-codesign-artifacts}
\end{figure*}

\subsubsection{Overview of what educators configured}
\label{sec:rq2-configuration-overview}

With an initial understanding of the GenAI experience through video probes and working with design materials, participants envisioned how an AI-supported student experience could address the challenges they had described. They made concrete what they wanted GenAI to do and what they wanted educators and students to control.  
% Their boards
% specified inputs such as student interests, teacher observations, and local
% resources. They also specified student-facing experiences such as career
% information, simulations, reflection, and next steps. Finally, they placed
% different forms of learner information between the activity and later human
% support, including summaries, interest records, alerts, and information that
% students carried between sessions.
Their configurations specified what students would receive or engage in, such as career information, simulations, reflection, and suggestions for next steps; what educators would supply or set, such as learning materials, local resources, and constraints on information sources; and when different forms of support would become available. 
Participants also considered how GenAI would interpret student activity, what it would remember across sessions, and what information would reach educators to inform subsequent guidance or intervention. They placed different forms of learner information between the activity and later human support, including summaries, interest records, alerts, and information that students carried between sessions.

Participants generally used the provided cards as prompts rather than assembling their designs from them. They typically began with one card and expanded the scenario using their own stickers and connections. Near the end, the researcher invited them to review the cards again; most selected at most one additional card. Across completed boards, participants created an average of approximately nine cards, suggesting that the cards primarily provided entry points for further elaboration.

The three examples in Figure~\ref{fig:illustrative-codesign-artifacts} show how these choices came together in different envisioned educational experiences. 
% T10 placed educator-selected sources before GenAI-generated
% activities. T14 connected student activity to a portable
% record that could later inform teacher planning. T15 connected a GenAI-generated
% interest profile to an inference timeline, privacy controls, and teacher
% redirection. 
T10 configured an experience in which educator-selected materials and source constraints shaped what GenAI offered students. T14 connected students' interactions with a portable record, allowing information from earlier activity to inform later teacher planning. T15 paired an AI-generated interest profile with a timeline of its inferences, controls over information visibility, and opportunities for a teacher to redirect the activity. 
These were not three versions of one workflow. They show that
participants configured different parts of the learning activity, from what
GenAI could draw on to how its output could reach a person. The following
sections examine these choices.

\subsubsection{Educators configured what GenAI personalized and how directly it guided students}
\label{sec:rq2-personalization-aspects}

A common pattern across participants' configurations was tailoring the GenAI-supported experience to students. Participants configured multiple aspects of what types of personalized support students can receive. 
%Participants configured personalization at several levels. 
Four participants
(T10, T13, T14, and T15) discussed adapting language, reading level, audio, or
other formats. T13, who teaches younger learners, noted that online career
information could be too technical and wanted GenAI to put it in
\miroquote{T13}{simpler terms that a kid might be able to understand}.

Participants also considered what information should inform this tailored support. 
T10 wanted to upload district materials and have GenAI
\miroquote{T10}{pull from my resources first, before they start going out to
other places}. In Figure~\ref{fig:illustrative-codesign-artifacts}, these source
choices define which materials are available before generation.  While T10 emphasized educators' role in selecting resources, T11 questioned how much adults should define the learner for whom support was personalized.   She wanted parents and
teachers to enter \miroquote{T11}{as little as possible} so adult assumptions
would not drive the system's suggestions. When considering whether an
adult might filter careers based on a disability, she stopped and said,
\miroquote{T11}{That might be too limiting}.
Together, these findings show that educators envisioned personalizing GenAI support through selected input resources and adaptations to student interactions, while questioning how far parents and teachers should shape that personalization through their own judgments about the learner.
% We interpret these configurations as distinguishing two uses of context in
% GenAI. Selected sources could ground a generated explanation or suggestion.
% Information about a learner could change the next generated response.
% Rewriting information in simpler language changed how a student accessed an
% option. Using learner or adult input to select careers could change which
% options were generated at all. The latter use could remove possibilities from
% view, as T11's concern about limiting careers illustrates.

\subsubsection{Educators configured when student activity became a GenAI inference}
\label{sec:rq2-identity-authorship}

RQ1 showed why educators valued observations gathered across activities and
settings. Participants envisioned using students' interactions with GenAI and the resulting records to inform subsequent support. However, when configuring the activity, participants highlighted that collecting data from student–AI interactions did not, by itself, provide a sufficient basis for interpreting students' interests. Five participants (T8, T10, T12,
T13, and T15) described or questioned this movement. T14 added that the
activity itself could yield weak evidence. If an interaction felt like a
standardized test, students might click through and produce
\miroquote{T14}{junk data}. %Thus, collecting more activity did not by itself
%make an inference more warranted in our analysis. 
 GenAI might treat students' clicks as evidence of interest even when those clicks did not reflect what interested them.

These concerns highlighted two configuration needs: making visible when GenAI turns activity records into inferences about students' interests, and providing opportunities to check those inferences with students.. T15 wanted the teacher view to show
\miroquote{T15}{a timeline of when the AI is making a decision about the
student's interest}. The timeline in Figure~\ref{fig:illustrative-codesign-artifacts}
records when the system updated an inferred interest as the student moved
across activities. T12 focused on checking an interpretation after it was formed. He
wanted to \premirofragment{do the double check} with the student. A mismatch between an
inferred interest in science and a student's stated interest in architecture
or art could \premiroquote{T12}{close the loop} by prompting clarification.

Together, these findings identify a need to configure how interpretations of student activity become visible and open to checking as they inform subsequent support. Participants sought ways to see when GenAI formed or updated an interest judgment and to involve students in assessing whether it reflected their interests.

\subsubsection{Educators configured whether learner information entered later GenAI sessions}
\label{sec:rq2-governed-continuity}

The previous two subsections examined what information should shape personalized support and how GenAI's interpretations of student activity could be made visible and checked. A further configuration question was whether and how these activity records and interpretations should carry forward to shape support in later sessions. 
Four of the 12 Miro boards explicitly addressed whether information should be retained for later sessions . T7 kept the supplied \miroquote{T7}{memory keeper} for
interests, choices, and earlier experiences. T13 rejected persistent memory. She connected
this choice to school data constraints, explaining,
\miroquote{T13}{If it's collecting and storing student data and remembering it
for the next time [\ldots] data collection and sharing data is a no-no in
education}.
T14 proposed a third arrangement: students could carry information between sessions without the system automatically retaining it.
 He suggested
\miroquote{T14}{a download [\ldots] with
everything that they thought} that the learner could upload in a later session.
He immediately noted that this extra step might be burdensome to students. In his case,
(Figure~\ref{fig:illustrative-codesign-artifacts}), continuity therefore depends
on a portable record and a learner's deliberate transfer, not hidden system
memory.

The contrast was not simply between memory and no memory. Participants
configured what persisted, whether it could be revised, and whether continuity
occurred automatically or through a learner's action.% We interpret these
% choices in terms of how earlier information enters later generation. T14's
% download-and-upload arrangement made reuse visible and deliberate. In
% contrast, automatic system memory could supply the same information to a later
% session without a comparable learner action.

\subsubsection{Educators configured what GenAI shared with people and what followed}
\label{sec:rq2-relational-loop}

Eight participants (T3, T4, T7, T9, T11, T12, T13, and T15) connected GenAI
activity to later human support. Their configurations addressed who should become involved, what information they would need, and when that involvement would be useful.  Seven proposed a GenAI summary or activity
report for educators. T13 wanted information that would help her
\miroquote{T13}{have a personal conversation with the kids about their
interests}. T9 similarly wanted a snapshot of the interaction.
It could become a \miroquote{T9}{conversation starter} and
\miroquote{T9}{help the teacher understand the child better because of this}.

Participants also placed boundaries on these connections according to the recipient's role. Six
(T3, T7, T12, T13, T14, and T15) raised privacy, institutional data rules, or
access for specific roles. T12 proposed that administrators see a summary of a class
or population \premiroquote{T12}{but not of the individual}. Individual
patterns could be available to a career counselor only when that person had
\premirofragment{a formal role in helping them select the career}. The same
learner information was not equally visible to every educator in T12's
proposal.

The information could support different next actions. T15 proposed that GenAI
\mirofragment{send a ping to the teacher} when a student
\miroquote{T15}{might be [\ldots] getting off track}. The teacher could then
interpret the situation and redirect or end the interaction. T11 extended the
connection beyond school staff. After digital exploration, students could
\miroquote{T11}{Take a tour of our company. Come, shadow us, come, do an
internship}. In her configuration, generated exploration led to a later
workplace experience with people. 
These configuration examples connected AI activity to different forms of human support, from intervention during an interaction to further exploration in a workplace.
% We interpret this sequence as separating
% career exposure from workplace experience.

% This finding differs from RQ1's concern with fragmented support. RQ1 explains
% why information from several activities or settings might be useful. RQ2 shows
% what educators configured once such information existed: its recipient, level
% of detail, trigger, and next action. A summary could prompt a conversation, an
% exception could prompt redirection, and a developed interest could prompt a
% connection to a workplace. We interpret sharing in these cases through the
% action the recipient was expected to take.

\subsubsection{Synthesis of the four findings}
\label{sec:rq2-boundary-matrix}

The opening RQ2 overview described what participants configured across their
boards. The next four subsections reported empirical patterns in those
configurations. We now reorganize the four findings into five possible configurations when educators design and configure for AI-supported activities:
what information GenAI uses to generate a response, when activity becomes an
inference, whether that inference enters a later session, who receives
generated information, and what happens next. Table~\ref{tab:boundary-matrix} summarizes these directions and the participant proposals that informed them.

\begin{table*}[t]
  \centering
  \caption{Five configuration directions identified across educators' envisioned GenAI-supported career-exploration experiences. The table relates participant configurations to our interpretation.
  Individual configurations were not necessarily shared by all participants.}
  \label{tab:boundary-matrix}
  \small
  \begin{tabularx}{\textwidth}{P{0.22\textwidth} P{0.43\textwidth} Y}
    \toprule
    Decision & What participants configured & Our interpretation \\
    \midrule
    \textbf{Information used to generate a response}
      & Participants prioritized selected materials, minimized adult input,
        and adapted language and format.
      & The configuration changed what GenAI could draw on and which
        possibilities a learner encountered. \\
    \textbf{Student activity used to infer an interest}
      & Participants questioned test-like evidence, added a timeline showing
        when GenAI inferred interests, and checked an inferred direction with the student.
      & Participants varied what evidence supported a claim, when GenAI made
        it, and who contributed the information. \\
    \textbf{Information reused in a later session}
      & Designs retained a memory, rejected memory, made a collection editable,
        or required the learner to transfer a record.
      & A present interpretation could, or could not, shape later activity. \\
    \textbf{Information shared with another person}
      & Participants configured summaries, access for specific roles, and privacy
        controls that limited access to selected information from an interaction.
      & Learner information crossed from the GenAI interaction to a particular
        person at a particular level of detail. \\
    \textbf{Information used for action}
      & Summaries prompted conversations, alerts prompted possible redirection,
        and developed interests prompted human or workplace experiences.
      & The information could affect guidance, intervention, or a future
        opportunity. \\
    \bottomrule
  \end{tabularx}
  \Description{Five rows show decisions about generating a response, inferring
  an interest, reusing information, sharing information, and taking action. Each
  row identifies participant configurations and our interpretation.}
\end{table*}

Each decision affected later uses without determining them.
An observation did not have to become an inference. An inference did not have
to persist. Persistent information did not have to be shared with every role,
and neither persistence nor sharing had to trigger action. Participants'
configurations made these changes conditional. We draw the following central
distinction across RQ2:
participants configured more than the content of a learner representation.
They also configured whether the information persisted, reached another
person, or shaped later output and action.

% These decisions provide the structure for RQ3. The first two concern what
% can shape GenAI output and when GenAI infers something about a learner. The next
% two concern whether learner information reaches a later session or another
% person. The final decision concerns what that information may affect.
\subsection{RQ3: Challenges and Support Needs in Envisioning GenAI Configurations}
\label{sec:design-synthesis}
RQ2 identified what educators wanted to configure in GenAI-supported career exploration. RQ3 examines the challenges and support needs that emerged as participants translated their teaching needs and practices into envisioned GenAI configurations. We identified three recurring challenges: moving beyond prevailing perceptions of AI toward concrete possibilities for educator control, breaking general-purpose GenAI into configurable functions and responsibilities, and identifying useful information from open-ended interactions based on intended teaching actions. Across these findings, we examine what made configuration difficult and what forms of support helped participants articulate AI support that could fit their teaching practice.

\subsubsection{Moving from prevailing perceptions of AI to configurable possibilities}
Participants could readily name teaching problems and broad goals, such as
saving time or providing individual support, as well as uses of AI they wanted
to avoid. Their concerns often reflected prevailing perceptions of GenAI,
particularly students using it to complete work without thinking for
themselves. T4 distinguished \premiroquote{T4}{using AI as a tool} from
\premiroquote{T4}{using AI as a crutch}, while T9 contrasted using ChatGPT to
produce a paper with its potential to \premiroquote{T9}{extend thinking}.
Participants therefore had ideas about what they wanted AI to support and what
they did not want it to replace, but these ideas did not readily specify what
an AI-supported activity should look like.

% Participants could name teaching problems and broad goals, such as saving time
% or providing individual support, but some struggled to turn them into new AI
% ideas. They could also articulate what they wanted to avoid, including students relying on AI to complete work without thinking for themselves. T4 distinguished \premiroquote{T4}{using AI as a tool} from \premiroquote{T4}{using AI as a crutch}, while T9 contrasted using ChatGPT to produce a paper with its potential to \premiroquote{T9}{extend thinking}.

Some participants struggled to translate these teaching needs and concerns
into concrete possibilities for GenAI configuration. 
However, knowing these goals and concerns did not mean knowing how to translate them into specific configurations. 
T5 could
identify her bottlenecks but explained,
\miroquote{T5}{I don't have enough AI knowledge to know [\ldots] the different
ways that AI could help me}. T9 likewise questioned whether teachers had the
professional development needed to \premiroquote{T9}{show them what's
possible}.
The challenge was therefore not only identifying teaching needs,
but understanding what could be configured to address them.

Participants' responses suggested that AI systems should make clear what educators could configure and how those choices could address their concerns. 
%, including filters and guardrails at different information-flow
%points very frequent and found that aspect most new and unfamilar to them.
They found the video probe is helpful in connecting their teaching language with the AI capabilities. Participants focused on possibilities for
shaping and constraining student-facing AI. T5 called the video
\premiroquote{T5}{a little bit mind-blowing} and later described it as
\premiroquote{T5}{really eye-opening [\ldots] to see how it can really be
[\ldots] protected and filtered}. T2 similarly said,
\premiroquote{T2}{I really like that the teachers can set up the exploration},
and emphasized repeated filtering \premiroquote{T2}{to make sure that it's
[\ldots] safe}. Making these forms of control concrete gave participants ways
to consider GenAI support beyond the uses they initially associated with it.
% These responses connected possible uses of AI with ways it
% might be constrained, rather than treating safeguards only as additions to an
%already defined use. 

These possibilities did not eliminate participants' concerns and invited further elaborations.
Participants also continued to question these
possibilities. T5 asked,
\premiroquote{T5}{Could it be broken? And
could my students be off and doing something that they shouldn't be?}.
% Their
% responses paired possible pedagogical uses with concerns about the safety and
% review those uses might require.
This points to a support need beyond
introducing educators to additional AI capabilities: educators need concrete
examples and opportunities to explore what they can configure and how those
choices shape students' experiences. Such support could help educators connect
their existing teaching needs and concerns to configuration possibilities
while making the consequences of different choices visible.

% What participants imagined was also bounded by familiar concerns about GenAI:
% that students could use it to cheat, avoid thinking, or access inappropriate
% or unreliable output. T1 described teaching students to use AI
% \premiroquote{T1}{in a real positive way [\ldots] not just to, like, cheat on
% things}. T4 distinguished \premiroquote{T4}{using AI as a tool} from
% \premiroquote{T4}{using AI as a crutch}.
% Participants considered possible forms of GenAI support alongside expectations
% that student-facing AI could be misused or unsafe.

\subsubsection{Turning general-purpose GenAI into configurable functions and responsibilities}

Even when participants could identify a pedagogical use, GenAI remained a
bundle of capabilities. The same interaction could generate an activity, ask
follow-up questions, adapt content, summarize a conversation, and infer
something about a learner. Treating these capabilities as one ``AI tool''
obscured decisions that educators wanted to make separately. Participants
instead broke the overall interaction into functions, considering what each
function would do, what information it would use, and what it would produce.

Considering individual functions involved more than identifying the features available. Participants also considered how different people would use those functions and how the functions would fit with people's participation in the learning activity. 
T5 noted, \miroquote{T5}{My responses so far have kind of been [\ldots] from
a classroom teacher [perspective]}, then said she would consider the system
differently from a counselor's perspective. A function therefore could not be
configured without also considering who would use its output and for what
responsibility.
Beyond identifying users, participants considered how AI functions would work alongside human activity. 

T8 said,
\premiroquote{T8}{I don't think AI should ever take over for a classroom
teacher, or the [\ldots] opportunity to actually be hands on}. T6 changed an
isolated student--AI exchange into \miroquote{T6}{a group type interaction
where the teacher might be interacting with the student and [\ldots] with AI}.

Most importantly, participants also considered how individual functions could influence one another. T8 described interests shaping generated scenarios, then
noted that students' choices within those scenarios could feed back into
inferred interests: \miroquote{T8}{that also kind of goes two ways}. % T6 raised

Configuring these functions therefore involved considering how a choice about one function could affect another. This suggests that educators need to understand both what individual functions do and how their connections shape the overall learning experience

% Participants similarly separated what AI could do from what people should
% continue to do. As reported in RQ2, they specified instructional goals, local
% courses, approved resources, grade expectations, and classroom rules as inputs
% to different functions.
% Configuration therefore included both assigning work to AI and preserving work
% for students and educators.
%a different boundary when immediate GenAI answers could replace students'
%productive struggle. 
% .Once these functions were separated, participants could
% specify not only what each function did, but when its output could travel,
% become evidence, or trigger another function---and where a person should remain
% involved.
% Across participants' configurations, we
% identified five: what AI does, what information it uses, who receives its
% output, who remains responsible, and whether its output becomes input to
% another function.
% Within the Miro activity, participants treated these decisions as separable. They could retain,
% remove, modify, and connect student-facing, teacher-facing, and supporting
% functions. Separating functions also exposed differences in who needed them.

\subsubsection{Identifying useful information from open-ended interaction through intended teaching actions}
Open-ended AI interactions can produce many forms of information, including
students' questions, choices, reflections, and patterns across interactions.
Participants did not treat all of this information as inherently useful for
teaching. Instead, they often identified what information should be surfaced
by considering what an educator would subsequently do with it. Intended
teaching actions therefore provided a practical basis for deciding what
information from an open-ended interaction was useful.
% The previous two themes show how educators translated teaching goals and
% concerns into configuration requirements and made sense of AI functions and
% their relationships. Participants also considered configuration from the other
% direction: what they would do after an AI-supported activity. Rather than
% treating more information about student interaction as inherently useful, they
% identified useful outputs in relation to existing or intended teaching actions
% and routines.

Educators determined what AI-generated information was useful based on what they would subsequently do with it. 
Seven of the 12 completed boards connected teacher input,
AI-supported student activity, and a later human response. T1's experience of working with about 170 students shaped his interest in reviewing \miroquote{T1}{what they're asking AI, and what AI is kind of giving to them}, which could help him identify students who needed \miroquote{T1}{the in-person communication}. T14 wanted GenAI to identify
\miroquote{T14}{common questions that students are having} because
\miroquote{T14}{I would put that into a lesson}. The intended teaching action
therefore helped determine which parts of an open-ended interaction were useful
to surface.
Across these cases, what educators intended to do next helped determine what
information from an open-ended AI interaction was useful to surface.

Information from students' interactions could also inform adjustments to the AI-supported activity itself. T15 considered adapting generated content based on students' selections. She linked what students did within an activity to how subsequent content could change.  Together, these findings suggest that supporting educators in configuring AI involves helping them understand how its use could shape their subsequent teaching, while opportunities to experience it in teaching practice could inform further configuration.
\section{Discussion}
\label{sec:discussion}
Open-ended learning can unfold across activities as learners' goals change and interpretations of their activity shape later support. Career exploration makes these dynamics particularly visible. Building on our findings, we discuss the scope and boundaries of GenAI configuration, how educators can be supported in expressing configuration needs, and the implications of GenAI-supported exploration for students' developing career interests and possibilities. We conclude by reflecting on our design method.

\subsection{Design Tradeoffs in Enabling Educator Configuration of GenAI-Supported Learning}
\label{sec:career-exploration-discussion}

As GenAI gets widely used in educational scenarios, GenAI systems increasingly shape students' learning experiences. This raises an important question for educators: how can these activities be incorporated into teaching and aligned with educational goals and contexts? Classroom orchestration research emphasizes educators' coordination of activities, resources, and participants so that different elements work together to support learning \cite{dillenbourg2013Orchestration,roschelle2013Orchestration}. GenAI-supported learning experiences likewise require educators' participation in their design, configuration, and adjustment.

Prior research provides important foundations for teachers to design AI activities, adjust system behavior, and review interactions \cite{ortegaArranz2025TeacherAutonomy,jin2025TeachTune}. We further ask which configuration choices matter for open-ended learning tasks, during which learners' goals evolve, the meaning of their activity emerges over time, and support spans multiple settings. Using career exploration as a case, our study identified five related configuration directions: personalizing student experiences, making interpretations visible and checkable, carrying information across activities, setting access to learner information, and connecting AI activities to human support (RQ2). These findings suggest that future systems for similar open-ended learning contexts should make these aspects available as configuration options for educators. Providing these options also requires making their scope and boundaries explicit. Below, we discuss several key boundaries and the tradeoffs involved in designing for them.

Adapting AI to students' needs was a recurring configuration direction. Prior research advocates for supporting teachers in adjusting AI responses through learning objectives, curricular materials, and student context to create personalized experiences for students \cite{ortegaArranz2025TeacherAutonomy}. Our findings further reveal that, in open-ended learning where goals are still developing, personalization concerns not only whether a response suits a student now, but also how the system characterizes that student and how this characterization affects later exploration. Future systems should therefore enable educators to specify both the scope and duration of personalization and make the effect transparent. For example, interfaces could show which content or options are affected by particular student information. Simulations could also allow educators to compare the support offered under different configurations and explore how reusing an early interpretation might change later activities \cite{lu2024generative}.

Providing connected support across long-term learning activities is an important concern for learner modeling and Agent memory research \cite{vanLabeke2007Interpretation}. Our findings echo this need for continuity while highlighting the distinction between retaining information and retaining an interpretation. Educators wanted to draw on students' past experiences to provide support, but did not want AI to overinterpret those experiences and gradually construct a rigid student profile. Open learner models have already emphasized opportunities to inspect and challenge systems' interpretations \cite{bullKay2016SMILI,vanLabeke2007Interpretation}. Configuring continued support additionally involves deciding what to remember, when to check again, stop using, or forget, and who has authority to make these decisions. One potential solution is to explicitly represent what is memorized by AI \cite{chen2026conversation} to help students and educators to configure it directly.

Our study also suggests that when GenAI activities connect to support from multiple parties, configurability should also involve distributing authority and responsibility among participants. Shared-control and classroom orchestration research emphasizes educators' guidance of and intervention in AI-supported activity \cite{lawrence2024SharedControl,holsteinOlsen2023CoOrchestration}. Our findings further suggest that teacher control should distinguish authority over learning activities from authority over learner information. In open-ended exploration, students should also be able to challenge interpretations of their interests and influence how these are shared and used. Systems could therefore combine role- and purpose-based access controls. The design goal is to align authority with educational responsibilities while preserving students' opportunities to express, question, and revise.

\subsection{Design Implications for Supporting Educators in Translating Teaching Needs into GenAI Configurations}

When educational AI tools are primarily shaped by system designers, educators have limited opportunities to configure how they support learning \cite{ortegaArranz2025TeacherAutonomy}. This can also limit their understanding of AI capabilities. In our study, educators' initial views of GenAI often centered on concerns about cheating, overreliance, and inappropriate content, while possibilities for configuring its behavior were less familiar to them. Our findings highlight the importance of involving educators in the design loop so they can explore how AI might be shaped around their educational goals. Yet such involvement requires support: knowing what students need does not necessarily mean knowing how to express those needs as AI configurations or determine what the system should produce. Through a design activity, our study explored how educators approached this translation and what support they needed to articulate their intentions. We draw on these findings to suggest how future GenAI systems could make configuration and control more accessible to educators.

First, configuration interfaces should help educators explore how their choices shape students' learning experiences. Related to teacher-driven personalization and pedagogical agent design \cite{walkington2025TeacherDriven,xu2026Explain,li2026GenRole}, our findings highlight the need to make unfamiliar configuration possibilities concrete. Prior work explored editable examples, interactive prototypes, and simulated interactions to help experts without a technical background author and explore AI experiences \cite{petridis2023PromptInfuser,jin2025TeachTune,kaputa2026SimStep}. Our work suggests future systems could link teacher settings with student-facing previews. This would help educators develop configuration ideas through seeing and testing their effects, including forms of support and control they may not initially know how to request.

Second, as AI systems combine more functions, understanding how they work together is more and more challenging for domain experts without technical backgrounds \cite{walkington2026MultiAgent, wu2022PromptChainer}. For example, recent educational systems distribute generation and evaluation across multiple agents \cite{walkington2026MultiAgent}. Our findings suggest that educator-facing interfaces should make these combinations understandable by decomposing them based on diverse educational purposes. Visual composition offers possible approaches \cite{wu2022PromptChainer,zhou2025InstructPipe}. Our findings suggest that systems should represent AI functions and their connections through language, visual representations, and concepts familiar to educators. For example, organizing them around teaching actions, student activities, and the information exchanged could help educators understand and configure their relationships without needing to master technical terms or architecture.

Finally, systems should support working backward from intended teaching actions to AI configurations. Building on learning analytics research linking information to classroom decisions \cite{verbert2014Dashboards,wiseJung2019TeachingAnalytics}, our findings suggest that showing how students might respond to AI-supported activities and how educators would follow up helped make their desired AI support more concrete. Systems could guide educators from these teaching purposes to the information needed and the activities that would produce it. Natural-language workflow authoring and incremental representations offer ways to support this translation \cite{zhou2025InstructPipe,kaputa2026SimStep}, while allowing educators to inspect and revise how their teaching intentions are implemented.

These implications may extend beyond career exploration to other longitudinal, evolving, and multi-stakeholder human–AI interactions, such as long-term project-based learning, AI companions for children, and AI-supported advising or planning. In these contexts, users' goals and needs develop over time, multiple people may participate in or act on AI-supported interactions, and information or interpretations from earlier interactions can shape later AI behavior and human action.

\subsection{Career Development in the Age of GenAI}
\label{sec:career-development-genai}

Our findings suggest that GenAI changes both the capabilities people need to
envision vocational futures and the relationships across which career
development occurs. We interpret these changes through Career Construction
Theory, possible selves, and the Systems Theory Framework.

\subsubsection{Critically Exploring AI-Generated Possible Selves}

Career Construction Theory associates career adaptability with resources that
include curiosity about possible futures and control over one's vocational
development \cite{savickas2005CareerConstruction}. Possible-selves research
similarly shows that adolescents try out, adjust, and abandon imagined futures
through activities and conversations with others
\cite{marshall2008PossibleSelves}. GenAI can extend this process by generating
occupations and experiences beyond those learners or their immediate social
worlds can readily name. This possibility aligns with educators' interest in
broadening exposure, but also with their resistance to narrowing exploration
or accelerating career decisions
(Sections~\ref{sec:rq1-open-exploration} and
\ref{sec:rq1-developmental-staging}).

These findings suggest that career adaptability in AI-mediated settings
requires more than receiving personalized possibilities. Learners must be able
to compare generated futures, examine the assumptions behind them, and decide
which to pursue, revise, or reject. This need is especially important because
GenAI may both present an experience and interpret the learner's response as
evidence of interest (Sections~\ref{sec:rq2-personalization-aspects} and
\ref{sec:rq2-identity-authorship}). Career education should therefore develop
curiosity through encounters with diverse futures while strengthening learners'
control over how AI-generated possibilities enter their evolving career
stories. This interpretation complements research on AI literacy and learner
agency by locating informed engagement with AI within career adaptability
\cite{longMagerko2020AILiteracy,vincoli2025Agency,retes2026Agency}.

\subsubsection{Career Development Across AI-Mediated Contexts}

The Systems Theory Framework understands career development as recursive and
shaped by interacting individual, social, educational, organizational, and
environmental influences \cite{pattonMcMahon2015Systems}. Participants likewise
located career learning across subjects, teachers, families, grade levels, and
community organizations, while describing observations as fragmented among
them (Section~\ref{sec:rq1-continuity}). GenAI may connect these influences and
sustain reflection over time. However, aggregation may separate an observation
from the activity or relationship that gave it meaning and may give a tentative
interpretation greater institutional authority.

Educators' contrasting approaches to persistence, learner-mediated transfer,
role-specific access, and human follow-up make this tension visible
(Sections~\ref{sec:rq2-governed-continuity} and
\ref{sec:rq2-relational-loop}). From a systems perspective, continuity should
not mean constructing an increasingly complete profile of a learner. Career
education should instead help students interpret multiple contextual
influences and sustain dialogue among students, teachers, families, and
counselors without treating their aggregation as a definitive account of who a
learner may become.

\subsection{Methodological Reflection on the Staged Design Probes}
\label{sec:phase-sensitive-analysis}

Design representations can make unfamiliar AI capabilities available for
ideation while also shaping what participants consider
\cite{yang2019SketchingNLP}. Technology and data probes similarly use concrete
artifacts to support engagement and critique
\cite{hutchinson2003TechnologyProbes,zhang2023DataProbes}, while participatory
artifacts can influence which possibilities become salient
\cite{sandersStappers2008Cocreation}. This tension was also evident in our iterative design process. Early open-ended sessions allowed educators to articulate teaching needs but provided limited support for imagining unfamiliar GenAI capabilities, while later storyboards made those capabilities more concrete but sometimes anchored participants to researcher-provided examples. We therefore iteratively developed the final protocol to balance concreteness with openness: the video established a shared understanding of possible GenAI roles and forms of educator control, while the reconfigurable Miro activity allowed participants to modify, reject, reconnect, or add to those possibilities. In the formal study, we further tracked where ideas became visible across these stages, distinguishing concerns expressed before the probes, responses to researcher-provided possibilities, and later elaborations of those possibilities.

We made this distinction in our analysis by examining each interview stage
separately and comparing completed boards with the starter materials. A
response to a demonstrated feature was treated as recognition, critique, or
elaboration rather than as a participant-originated requirement. Retaining a
starter card indicated use of a provided possibility, while modifying,
rejecting, or reconnecting it showed how participants changed or extended that
possibility. Appendix Table~\ref{tab:finding-provenance} traces where the
central findings first became visible, what the study materials had already
introduced, and how participants changed or extended them.

Coordination Theory informed the design of the Miro probe
\cite{maloneCrowston1994Coordination}. We represented the proposed system
through actors, activities, and dependencies, using panels, function cards,
and connectors that participants could arrange and modify. This representation
gave participants a way to discuss who performed an activity, what passed
between activities, and where responsibilities or handoffs occurred without
requiring them to describe a technical architecture. The same concepts gave us
a vocabulary for describing the function and dependency decomposition reported
in RQ3.

Tracking ideas across stages helps preserve the relationship between what the study introduced and what participants contributed within this co-constructed design activity. The probes necessarily shaped parts of the design space participants considered by making particular AI functions and relationships concrete and available for configuration. Our findings therefore should not be interpreted as identifying an exhaustive or optimal decomposition of GenAI-supported learning, nor as showing what educators would have designed without these representations. Instead, our analysis focuses on how participants worked within and beyond the provided design space: some concerns appeared before the probes, while participants later rejected, modified, and reconnected introduced possibilities, such as removing persistent memory or specifying conditions for handoffs. This makes visible how domain experts reason about and configure unfamiliar AI functions once those possibilities become tangible, while recognizing that alternative representations may surface different configurations.

\subsection{Limitations}

Career exploration was a useful setting because it is open-ended, develops
over time, and involves students, educators, counselors, families, and local
institutions. These features made questions about inference, memory, sharing,
and human follow-up easier to see. They also limit how broadly the findings can
be applied. Tasks with stable answers, short time frames, or fewer people may
involve different configuration needs. Our design implications may therefore be
most relevant to settings in which learner goals develop over time and AI
interpretations can shape subsequent support or opportunities.

The sample was small and regionally concentrated, and it included classroom
teachers, educational leaders, a counselor, workforce-development personnel,
and homeschooling parents. These roles carry different responsibilities, but
the study was not designed for systematic comparison among them. Student and
family perspectives were also limited: students did not participate, and the
family perspective came from the participating homeschooling parents. Prior
work shows that educators and students can differ in how they evaluate AI
explanations, over-reliance, and control
\cite{leeSong2024Explanations,chanTsi2024Perceptions,yang2026BalancingAgency}.
The learner-facing controls in our findings therefore represent participating
adults' proposals, not evidence of what students would understand, prefer, or
consider fair.

Finally, the study examined responses to a video probe and proposed
configurations and did not include the use of an implemented system. The implications of
this elicitation process are discussed in
Section~\ref{sec:phase-sensitive-analysis} and are not repeated here. Not every
interview produced a completed Miro board, so artifact-based claims draw only on
the available boards. The study did not assess technical feasibility, learning
outcomes, privacy consequences, educator workload, or sustained use. Future
work could involve students and families as design partners and evaluate these
configurations through longer-term use in schools.

\section{Conclusion}

We studied how 15 educators envisioned configuring GenAI-supported career
exploration across learning activities and human guidance. Participants
configured GenAI functions, information sources and flows, GenAI inferences
about learners, persistence, disclosure, and handoffs to people. They grounded these choices in
pedagogical purposes, divisions of responsibility, and intended teaching
actions. These findings motivate two design directions for open-ended
GenAI-supported learning. Educator-facing orchestration could help educators
author role-bounded functions, inspect their dependencies, and connect outputs
to human action. Separate controls over inference, persistence, disclosure, and
action could keep learner representations revisable and clarify who may use
them for educational decisions. These directions do not prescribe a particular
technical architecture. They identify interface and system controls that may
help connect GenAI support to educational practice while learners' goals and
needs continue to develop.

%%
%% The next two lines define the bibliography style to be used, and
%% the bibliography file.
\bibliographystyle{ACM-Reference-Format}
\bibliography{references}

%%
%% If your work has an appendix, this is the place to put it.
\appendix
\clearpage
\onecolumn

\section{Supplementary Findings Tables}
\label{app:findings-tables}

\section{Guide to Analysis by Interview Stage}
\label{app:phase-analysis}

\begin{table*}[h]
  \centering
  \caption{Analytic distinctions used across the staged protocol.}
  \label{tab:phase-guide}
  \small
  \begin{tabularx}{\textwidth}{P{0.16\textwidth} P{0.22\textwidth} Y Y}
    \toprule
    Stage & Primary material & Analytic focus & Claim limitation \\
    \midrule
    Unaided interview & Interview questions about current practice
      & Problems, routines, constraints, and existing human work
      & Does not show how a participant would configure an unfamiliar AI system. \\
    Response to video probe & Participant response to the researcher-authored five-minute demonstration
      & What participants noticed, questioned, and connected to their work
      & Endorsement of a shown feature is not independent demand. \\
    Miro design artifact and think-aloud & Completed Miro board and the participant's think-aloud explanation
      & Who performs each function, how information flows, where control lies,
        and when handoffs or stopping points occur
      & Participants and researchers co-created the boards. The boards do not show implemented behavior. \\
    Final reflection & Participant comments on the completed design and study materials
      & How participants understood their configurations and how the probes supported or constrained articulation
      & Retrospective comments do not establish that the probes caused a change in participants' thinking. \\
    \bottomrule
  \end{tabularx}
\end{table*}

\begin{table*}[p]
  \centering
  \caption{Provenance of the four central RQ2 findings across the staged study.
  Earliest phase means the first recorded evidence in this dataset, not when an
  idea originated or what caused it. The final column reports the
  transformations most consequential to each finding. It does not count every
  board edit.}
  \label{tab:finding-provenance}
  \small
  \begin{tabularx}{\textwidth}{P{0.19\textwidth} P{0.13\textwidth} P{0.25\textwidth} Y}
    \toprule
    Central finding & Earliest phase in the dataset & Related material supplied by the researchers & How participants changed or extended it \\
    \midrule
    \textbf{Educators configured what GenAI personalized and how directly it guided students}
      & Unaided interview
      & The video showed several interaction formats and teacher authoring. The Miro board supplied Activity Designer, Career Simulator, Reflection Partner, and Next-Step Guide cards.
      & T10 gave priority to district sources. T11 minimized information from adults to limit its effect on generated suggestions. T13 configured language for younger learners. \\
    \textbf{Educators configured when student activity became a GenAI inference}
      & Video response
      & The video showed a dashboard with a GenAI-generated interest profile. The Miro example supplied retained context, but not an inference timeline, student confirmation, or a record of who contributed information.
      & T14 rejected test-like interaction as weak evidence. T15 added a timeline of GenAI interest decisions. T12 checked an inferred direction with the student. \\
    \textbf{Educators configured whether learner information entered later GenAI sessions}
      & Unaided interview
      & The Miro board supplied a Memory Keeper and illustrated selected reflections being retained and shared with permission.
      & T13 rejected persistent memory. T14 asked learners to download information and upload it in a later session. T11 made the collection revisable and student driven. \\
    \textbf{Educators configured what GenAI shared with people and what followed}
      & Video response
      & The video showed teacher dashboards and prompted discussion of educator involvement. The Miro example supplied Teacher Handoff and sharing with permission.
      & Participants specified the trigger, information, recipient, and next action. T3 connected a checked summary to human discussion. T13 connected selected activity signals to a teacher conversation. T15 added privacy alerts and live redirection. T11 extended handoff to employers and authentic experiences. \\
    \bottomrule
  \end{tabularx}
  \Description{Four rows identify when each central RQ2 finding was first
  visible, which related video or Miro elements the researchers supplied, and
  how participants changed, rejected, or extended those elements.}
\end{table*}

\clearpage
\onecolumn
\section{Full Analytic Codebook}
\label{app:codebook}

Table~\ref{tab:full-codebook} lists the final thematic codebook used to report
patterns across participants. Section~\ref{sec:data-analysis} describes the
artifact-level codes used to analyze the Miro boards. Quotations are kept in
the Findings and are not repeated here. This choice keeps each excerpt in one
place while providing the thematic code definitions.

\begingroup
\setlength{\tabcolsep}{5pt}
\renewcommand{\arraystretch}{1.12}
\scriptsize
\begin{longtable}{@{}P{0.35\textwidth}P{0.61\textwidth}@{}}
\caption{Full analytic codebook grouped by theme.}
\label{tab:full-codebook}\\
\toprule
\textbf{Code} & \textbf{Definition} \\
\midrule
\endfirsthead
\multicolumn{2}{l}{\tablename~\thetable, continued}\\
\toprule
\textbf{Code} & \textbf{Definition} \\
\midrule
\endhead
\midrule
\multicolumn{2}{r}{Continued on next page}\\
\endfoot
\bottomrule
\endlastfoot

\multicolumn{2}{@{}l}{\textbf{Developmental career support}} \\
\texttt{CAREER\_AWARENESS\_GAP} & Student has limited knowledge of careers available to explore. \\
\texttt{PATHWAY\_TRANSLATION} & AI connects an expressed interest to education and career routes. \\
\texttt{AUTHENTIC\_OCCUPATIONAL\_EXPERIENCE} & AI makes everyday career practices easier to encounter. \\
\texttt{REFLECTIVE\_INTERPRETATION} & AI or educators help students make sense of experiences while leaving room for uncertainty. \\
\texttt{EXPLORATION\_TO\_EXECUTION} & AI connects exploration to concrete education steps. \\
\texttt{EVOLVING\_TRAJECTORY} & Support follows changes in plans, interests, and experiences over time. \\
\texttt{DEVELOPMENTALLY\_STAGED\_SUPPORT} & Career support moves from early exploration to later pathway and credential action. \\
\texttt{NONCLOSING\_EXPLORATION} & Early career learning builds awareness without fixing a student to a predicted career. \\

\addlinespace
\multicolumn{2}{@{}l}{\textbf{Implementation context}} \\
\texttt{SCALE\_AND\_TIME\_PRESSURE} & Large caseloads and limited time constrain individual support. \\
\texttt{SCHOOL\_ROUTINE\_INTEGRATION} & AI-supported work fits advisory, curriculum, grade level, and planning routines. \\
\texttt{FRAGMENTED\_RESOURCE\_CONSOLIDATION} & AI connects existing assessments, websites, data, and activities. \\
\texttt{TEACHER\_AI\_LITERACY\_AND\_OVERWHELM} & Educators recognize needs but may struggle to express them as AI roles or settings. \\

\addlinespace
\multicolumn{2}{@{}l}{\textbf{Institutional capacity and workflow}} \\
\texttt{LOCAL\_LABOR\_MARKET\_GROUNDING} & Guidance connects national career data to nearby jobs and employers. \\
\texttt{FAMILY\_GUIDANCE\_INEQUITY} & Students receive unequal pathway support because families hold different education and career knowledge. \\
\texttt{DISTRICT\_RESOURCE\_PRIORITY} & AI checks approved district and local resources before broader sources. \\

\addlinespace
\multicolumn{2}{@{}l}{\textbf{Human--AI boundary work}} \\
\texttt{HUMAN\_CONTEXTUAL\_INTERPRETATION} & People retain judgment involving values, family, identity, and local effects. \\
\texttt{FACTUAL\_BREADTH\_AI} & AI provides broad factual and career information. \\
\texttt{HUMAN\_RELATIONSHIP\_NONREPLACEMENT} & AI strengthens face-to-face interaction and leaves relational work to people. \\
\texttt{TRIADIC\_COLLABORATION} & Student, educator, and AI take part in a shared interaction. \\
\texttt{TEACHER\_COUNSELOR\_DIFFERENTIATION} & Teachers and counselors need different information, functions, and interfaces. \\

\addlinespace
\multicolumn{2}{@{}l}{\textbf{Coordination and handoffs}} \\
\texttt{AI\_PREPARES\_FOCUSED\_DEBRIEF} & AI-supported exploration prepares a later one-to-one conversation. \\
\texttt{LAUNCHING\_POINT\_FOR\_HUMAN\_INTERACTION} & AI output starts a human conversation and leaves the issue open for discussion. \\
\texttt{ALERT\_AND\_TRIAGE} & AI brings time-sensitive wellbeing or safety concerns to the educator assigned by school policy to respond. \\
\texttt{PERSONALIZATION\_DECISION\_TIMELINE} & Educators can review the sequence of AI-directed changes or interest updates. \\
\texttt{COORDINATOR\_AS\_BOUNDARY\_MANAGER} & A coordination function routes needed context, timing, permission, and responsibility. \\
\texttt{AI\_TO\_EDUCATOR\_SUMMARY} & AI summarizes student or class activity for educator action. \\
\texttt{TEACHER\_AI\_TO\_STUDENT\_AI} & Teacher-facing settings or instructions change student-facing AI behavior. \\

\addlinespace
\multicolumn{2}{@{}l}{\textbf{Personalization and agency}} \\
\texttt{CONTESTABLE\_PROFILE} & Students can question whether an AI interpretation is accurate or acceptable. \\
\texttt{STATED\_INTEREST\_RESPONSIVENESS} & The system responds to interests stated by the student, including unexpected careers. \\
\texttt{MODALITY\_CHOICE} & Students choose among chat, reading, game, and other formats. \\
\texttt{LOCAL\_PATHWAY\_GROUNDING} & Personalization uses courses, credits, and pathways that are available locally. \\
\texttt{ADAPTIVE\_SCAFFOLDING} & Difficulty, background support, or enrichment changes with student needs. \\
\texttt{ACCESSIBLE\_LANGUAGE\_AND\_READING\_LEVEL} & Career content is translated or rewritten for language and reading level. \\
\texttt{USER\_MEDIATED\_PORTABLE\_MEMORY} & A learner deliberately transfers a limited record to a later session. \\

\addlinespace
\multicolumn{2}{@{}l}{\textbf{AI-generated experiences}} \\
\texttt{EXPERIENCE\_TO\_REFLECTION} & Activity behavior becomes tentative material for reflection. \\
\texttt{GAMIFIED\_ENGAGEMENT} & Game-like interaction supports initial engagement with career learning. \\
\texttt{TEACHER\_PREVIEW\_AND\_AGE\_FIT} & Teachers preview generated material and check whether it fits the age group. \\
\texttt{REAL\_WORLD\_EXPERIENCE\_BOUNDARY} & Simulations support preparation while hands-on experience remains important. \\
\texttt{ENGAGEMENT\_AUTHENTICITY\_TENSION} & An activity balances career learning with genuinely engaging play. \\
\texttt{LONG\_TAIL\_GENERATION} & Generation supports careers beyond a fixed content library. \\
\texttt{EXPERIENCE\_TO\_TEACHER\_PLANNING} & Patterns from student experiences inform teacher planning. \\
\texttt{TESTLIKE\_INTERACTION\_TO\_JUNK\_EVIDENCE} & Test-like interaction can produce traces that poorly represent interest or engagement. \\

\addlinespace
\multicolumn{2}{@{}l}{\textbf{Boundary-crossing mechanisms}} \\
\texttt{OBSERVATION\_TO\_INFERENCE\_SLIPPAGE} & AI turns an action or response into an inference with weak evidence. \\
\texttt{INDIRECT\_SIGNAL\_ACCUMULATION\_AND\_CONFIRMATION} & AI keeps weak signals tentative until the student confirms or corrects them. \\
\texttt{INFERENCE\_TO\_IDENTITY\_CLOSURE} & AI presents a tentative inference as a stable description of the student. \\
\texttt{AI\_BORROWED\_HUMAN\_CONTROL} & AI presents an interpretation as if it were a teacher's or counselor's decision. \\
\texttt{SUPPORT\_TO\_DECISION\_SUBSTITUTION} & AI moves from providing support to making a high-impact decision that needs human judgment. \\
\texttt{SCAFFOLDING\_TO\_ANSWER\_REPLACEMENT} & AI supplies answers in ways that reduce productive struggle and reflection. \\
\texttt{SIMULATION\_TO\_REAL\_WORLD\_SUBSTITUTION} & An AI simulation is treated as equal to hands-on, social, or workplace experience. \\
\texttt{RAW\_ACTIVITY\_TO\_INSTITUTIONAL\_DISCLOSURE} & Student activity becomes visible to an institution without a clear purpose or limit. \\
\texttt{VISIBILITY\_TO\_SURVEILLANCE} & A support tool becomes continuous monitoring or broad exposure. \\
\texttt{PREMATURE\_OR\_DELAYED\_HANDOFF} & Human support enters too early for exploration or too late to affect an outcome. \\

\addlinespace
\multicolumn{2}{@{}l}{\textbf{Trust and safety}} \\
\texttt{PRIVACY\_SURVEILLANCE\_TENSION} & Educator visibility or safety monitoring may invade student privacy. \\
\texttt{GUARDRAIL\_RELIABILITY} & Educators value safety filters but question whether they can fail. \\
\texttt{STUDENT\_AI\_USE\_NORMS} & Students learn the difference between using AI as a tool and relying on its answers. \\

\addlinespace
\multicolumn{2}{@{}l}{\textbf{Teacher governance}} \\
\texttt{TEACHER\_VISIBILITY} & Teachers receive relevant information about student exploration. \\
\texttt{PII\_INPUT\_PROTECTION} & The system discourages, detects, or blocks personal identifying information. \\
\texttt{INSTITUTIONAL\_DATA\_GOVERNANCE} & School approval depends on how a platform collects, stores, remembers, and shares data. \\
\texttt{ACTIVE\_EDUCATOR\_INVOLVEMENT} & Visibility allows educators to join, guide, and plan. \\
\texttt{LOCAL\_CONFIGURATION\_AND\_CURATION} & Educators provide local knowledge, adjust content, preview output, and set safeguards. \\
\texttt{LIVE\_REDIRECTION\_CONTROL} & An educator can redirect, pause, message, or change an off-task experience. \\

\addlinespace
\multicolumn{2}{@{}l}{\textbf{Engagement with the method}} \\
\texttt{STUDENT\_DATA\_CONTROL\_GAP} & Educators gave limited detail about student control over memory, sharing, and AI settings. \\
\texttt{FEATURE\_RECOGNITION\_AND\_RESONANCE} & The video helped educators recognize capabilities and connect them to current problems. \\
\texttt{ROLE\_ALLOCATION\_AND\_RELATIONAL\_ELABORATION} & The design activity helped participants place functions and explain relationships and handoffs. \\
\texttt{AGENT\_TERMINOLOGY\_AND\_AI\_LITERACY\_BURDEN} & Participants knew work problems but struggled to express them as AI roles. \\
\texttt{VIDEO\_OPENNESS\_VERSUS\_CONCRETENESS} & A concrete video supports discussion but can anchor attention to features it shows. \\
\texttt{PRE\_VIDEO\_EXPECTATION\_GAP} & The protocol did not fully capture expectations just before the video. \\
\texttt{COLLABORATIVE\_SCHOOL\_ENVISIONING} & Educators work together to place a proposed design within school routines. \\

\addlinespace
\multicolumn{2}{@{}l}{\textbf{Multi-actor boundary configuration}} \\
\texttt{RELATIONAL\_LOOP\_CLOSURE} & AI-supported activity returns needed information to a person who can take the next action. \\
\texttt{EXCEPTION\_BASED\_ORCHESTRATION} & AI identifies a possible need so educators can focus limited attention and judge the context. \\
\texttt{AI\_SESSION\_BOUNDARY\_CONTROL} & The teacher responsible for the activity decides when AI begins, pauses, redirects, hands off, or ends. \\
\texttt{TEACHER\_SEEDS\_LEARNER\_CONTEXT} & An educator provides limited context that guides AI personalization. \\
\texttt{TEACHER\_GOVERNED\_EXTERNAL\_RETRIEVAL} & Educators set local sources and filtering rules for external retrieval. \\

\end{longtable}
\endgroup

\end{document}